\documentclass[twocolumn,trackchanges]{aastex701}

\usepackage{hyperref}
\usepackage{url}
\usepackage{caption}
\usepackage{adjustbox,booktabs}
\usepackage{amsmath}
\begin{document}

\title{Unsupervised Machine Learning Classification of FRB 20121102A across Distinct Active Episodes: Implications for FRB Emission physics 
and Source Environments}

\author[0009-0001-5037-3855,sname='Li']{Xiao Li}
\affiliation{Guangxi Key Laboratory for Relativistic Astrophysics, School of Physical Science and Technology, Guangxi University, Nanning 530004, China}
\email[show]{lxgo@st.gxu.edu.cn}  

\author[0000-0002-7044-733X]{En-Wei Liang}
\affiliation{Guangxi Key Laboratory for Relativistic Astrophysics, School of Physical Science and Technology, Guangxi University, Nanning 530004, China}
\email[show]{lew@gxu.edu.cn}
\correspondingauthor{En-Wei Liang}

\begin{abstract}
Fast radio bursts (FRBs) are millisecond-duration radio transients of extragalactic origin. Identifying distinct burst populations and investigating their variations across active episodes provide important insights into the physical origins of repeating FRBs.
In this paper, we apply an unsupervised machine learning approach to bursts from three active episodes of FRB 20121102A observed by the Five-hundred-metre Aperture Spherical Telescope (FAST). 
Our analysis reveals four distinct clusters primarily
distinguished by their spectral properties.
Notably, Cluster 1 exhibits bimodal-like distributions in central frequency and bandwidth and contains nearly all high-energy bursts, with a mean energy about an order of magnitude higher than those of the other clusters.
It also occupies a distinct region in the width--fluence parameter space. These results suggest that this cluster may represent a distinct burst population potentially associated with different emission physics.
Furthermore, the fractions of different burst clusters vary across active episodes. Cluster 1 accounts for about 27\% of the bursts in 2019, compared with about 8\% in 2022 and 2023. Cluster 2, with higher central frequencies and narrower bandwidths, has the largest fraction in 2019 and 2022, whereas Cluster 3, with broader bandwidths and intermediate central frequencies, has the largest fraction in 2023.
These variations may indicate changes in the physical conditions of the emission regions or the local propagation environment across active episodes.
\end{abstract}


\section{Introduction}
Fast radio bursts (FRBs) are bright, short-duration radio transients characterized by dispersion measures (DMs) significantly exceeding the expected Galactic contribution, indicating their extragalactic origin \citep{2007Sci...318..777L,2013Sci...341...53T,2019ARA&A..57..417C,2022A&ARv..30....2P,2023RvMP...95c5005Z}.  
Despite extensive observational efforts, their physical origin remains unclear and is still under active debate (see \citealt{2019PhR...821....1P,2023RvMP...95c5005Z} for reviews).
The association between the Galactic magnetar SGR 1935+2154 and FRB 20200428 suggests that at least some FRBs may originate from magnetars \citep{2020Natur.587...59B,2020Natur.587...54C}. 
FRBs are commonly classified into repeaters and non-repeaters based on their observed burst activity.
To date, more than 4000 FRBs have been detected, among which over 80 have been identified as repeaters \citep{2021ApJS..257...59C,2024ApJ...969..145C,2026ApJS..283...34C}\footnote{\url{https://blinkverse.zero2x.org/overview}}. 
A subset of repeaters has exhibited hyperactivity, with thousands to tens of thousands of bursts already detected (e.g., \citealt{2021Natur.598..267L,2022Natur.606..873N,2022Natur.609..685X,2023ApJ...955..142Z,2025arXiv250714707Z,2026arXiv260216409U}). 

Some hyperactive repeaters exhibit burst activity concentrated in distinct active episodes lasting from several days to several months \citep{2021Natur.598..267L,2023MNRAS.519..666J,2025arXiv250715790W,2022Natur.609..685X,2022RAA....22l4001Z}. Among them, FRB 20121102A is one of the most extensively studied FRB sources.
It was the first repeater to be discovered and was subsequently localized to a star-forming region in a dwarf galaxy at $z \sim 0.193$ \citep{2014ApJ...790..101S,2016Natur.531..202S,2017Natur.541...58C}.
Several active episodes have been observed by multiple radio telescopes worldwide (e.g., \citealt{2018ApJ...863....2G,2018ApJ...866..149Z,2019A&A...623A..42H,2019ApJ...882L..18J,2020A&A...635A..61O,2020MNRAS.495.3551R,2021Natur.598..267L,2022MNRAS.515.3577H,2023MNRAS.519..666J,2024MNRAS.527.9872G,2025A&A...693A..40B,2025A&A...704A..25G,2025arXiv250715790W}), including three major episodes monitored by the Five-hundred-meter Aperture Spherical radio Telescope (FAST) in 2019, 2022, and 2023. 
Notably, significant variations in burst properties have been observed among these different episodes.
During the 2019 active episode, the source displayed a bimodal energy distribution, whereas this bimodality was not observed during the 2022 and 2023 active episodes. Furthermore, even within a single active episode, individual bursts show substantial diversity in energy, flux density, central frequency, and bandwidth \citep{2021Natur.598..267L,2025arXiv250715790W}. 
These observational results suggest that bursts from FRB 20121102A may comprise multiple sub-populations, potentially associated with different radiation mechanisms or distinct physical conditions in the emission regions \citep{2021Natur.598..267L,2022A&A...657L...7X}. 
Therefore, identifying distinct burst populations and investigating their variations across active episodes are essential for understanding the physical origins of active repeaters.

Machine learning has been widely applied to the classification of FRBs, with studies mainly divided into two categories. The first category applies machine learning to data from the Canadian Hydrogen Intensity Mapping Experiment to identify repeater candidates among sources initially classified as non-repeaters (e.g., \citealt{2022MNRAS.509.1227C,2023MNRAS.518.1629L,2023MNRAS.519.1823Z,2023MNRAS.522.4342Y,2026ApJ..1000..181T,2026MNRAS.545f2178M,2026ApJ...998..154K,2026SCPMA..6909512S}).
In particular, \citealt{2022MNRAS.509.1227C} and \citealt{2023MNRAS.519.1823Z} applied unsupervised machine learning methods to the first CHIME/FRB catalog and identified 188 and 117 repeater candidates, respectively, among 474 sources initially classified as non-repeaters. 
Follow-up observations confirmed that some of these candidates are repeaters \citep{2023ApJ...947...83C,2025ApJ...982...16Q,2025ApJ...980..185S}, demonstrating the potential of machine learning for FRB classification. The second category focuses on applying machine learning to large samples of bursts from highly active repeaters to investigate burst diversity and possible underlying physical origins (e.g., \citealt{2023MNRAS.524.1668R,2023MNRAS.521.5738C,2024PASA...41...90L,2026PASP..138b4101H,2026A&A...713A.179Z}). For example, \cite{2023MNRAS.521.5738C} applied unsupervised machine learning techniques to a burst sample from FRB 20201124A obtained by FAST during an active episode and identified three burst clusters with distinct physical properties, suggesting the existence of multiple burst sub-populations.

Machine learning techniques provide a powerful and data-driven framework for simultaneously analyzing multidimensional burst properties and identifying distinct populations within repeating FRBs. In this paper, we apply an unsupervised machine learning approach to bursts from FRB 20121102A detected by FAST during three active episodes and identify distinct burst clusters. By comparing the distributions and properties of these clusters across different active episodes, we investigate whether the burst populations vary with source activity and explore their possible physical origins.
This paper is structured as follows: Section \ref{sec2} details the data and methods. Section \ref{sec3} presents the results of the analysis. 
Discussions and conclusions are presented in Sections \ref{sec4} and \ref{sec5}, respectively.

\section{Data and Methods}\label{sec2}
\subsection{Sample Construction and Feature Selection}
Our analysis is based on three burst samples from FRB 20121102A observed by FAST during different active episodes.  
During the active episode in 2019, a sample of 1652 bursts from this source was detected with the FAST telescope \citep{2021Natur.598..267L}. 
Subsequent FAST monitoring campaigns detected 431 and 124 bursts during the 2022 and 2023 active episodes, respectively \citep{2025arXiv250715790W}. All three observations cover the $1000-1500$ MHz frequency range and employ similar single-pulse search procedures.  To reduce potential selection biases associated with the different search procedures, we impose a common lower energy threshold of $E \geq 2.5\times10^{37}\ {\rm erg}$ on all three samples. This corresponds to the 90\% detection completeness threshold estimated by \cite{2021Natur.598..267L} for bursts with a characteristic width of 3 ms and was also adopted by \cite{2025arXiv250715790W} in their analysis. Note that the 1652 bursts reported by \cite{2021Natur.598..267L} do not include measurements of the central frequency. We therefore adopt the central frequencies and bandwidths measured by \cite{2025arXiv251007002W} for the same 1652 bursts in their reanalysis of the FAST L-band archival observations. The derived energy, width, fluence, and peak flux measurements are consistent with those reported by \cite{2021Natur.598..267L}.

Each burst in the three samples includes the following observational parameters: burst arrival time (MJD), dispersion measure (pc\,cm$^{-3}$),  width (ms), bandwidth (MHz), central frequency (MHz), peak flux (mJy), fluence (mJy\,ms), and energy (erg). To enrich the feature space and improve the robustness of our analysis, we introduce two derived parameters: the bandwidth ratio, defined as the ratio of bandwidth to central frequency and commonly used to characterize FRB spectral properties \citep{2023ApJ...956...67Y,2024ApJ...974..160K}, and the waiting time, defined as the time interval between consecutive bursts and used in previous studies \citep{2023MNRAS.524.1668R,2023MNRAS.521.5738C}. Note that we exclude bursts with waiting times greater than or equal to one day, since such long intervals are likely caused by observational gaps in the monitoring campaign rather than the intrinsic activity of the source.

For feature selection, we exclude the burst arrival time, as the long observational gaps between different active episodes may introduce artificial structures into the clustering results. We also exclude the dispersion measure, since it primarily traces the integrated electron column density along the propagation path and may introduce propagation-related effects that are not directly associated with the intrinsic burst properties. In our analysis, all bursts are required to have valid values for the eight selected features: width, bandwidth, central frequency, bandwidth ratio, peak flux, fluence, energy, and waiting time.
Adopting the above selection criteria, we analyze the bursts from the three active episodes together as a single dataset. The final sample used for the unsupervised machine learning analysis consists of 2006 bursts in total, with 1501, 390, and 115 bursts from the 2019, 2022, and 2023 active episodes, respectively. These selected features are then used as inputs for the dimensionality reduction and clustering analyses described in the following sections.

\subsection{Unsupervised machine learning techniques}
Unsupervised machine learning is a powerful approach for uncovering hidden structures in data without prior classification information \citep{FOTOPOULOU2024100851,2025Univ...11..412K}.   
Dimensionality reduction and clustering are two widely used methods in unsupervised machine learning. Dimensionality reduction projects high-dimensional data into a lower-dimensional space, allowing visualization and interpretation of the underlying data structure. Clustering algorithms then identify clusters of data points with similar characteristics in the embedding space. 
In this paper, we employ Uniform Manifold Approximation and Projection\footnote{\url{https://umap-learn.readthedocs.io/en/latest/index.html}} (UMAP; \citealt{2018arXiv180203426M}) for dimensionality reduction, followed by Hierarchical Density-Based Spatial Clustering of Applications with Noise\footnote{\url{https://hdbscan.readthedocs.io/en/latest/index.html}} (HDBSCAN; \citealt{10.1007/978-3-642-37456-2_14,Campello2015HierarchicalDE,2017JOSS....2..205M}) to identify clusters in the resulting low-dimensional embedding space.

\subsubsection{Uniform Manifold Approximation and Projection}
UMAP is a nonlinear dimensionality reduction technique based on manifold learning, with its theoretical foundation rooted in Riemannian geometry and algebraic topology \citep{2018arXiv180203426M}. It assumes that the data are uniformly distributed on a locally connected Riemannian manifold whose metric is locally constant or can be approximated as such. Under these assumptions, UMAP constructs a fuzzy topological representation of the high-dimensional data to capture its intrinsic geometric properties. A low-dimensional embedding is then obtained by optimizing the correspondence between the high- and low-dimensional representations, preserving local relationships while retaining as much of the global structure as possible. Therefore, UMAP is particularly useful for visualizing and exploring high-dimensional datasets in a low-dimensional space.

UMAP has several hyperparameters that can have a significant impact on the resulting embedding. Here, we briefly describe three key parameters: \texttt{n\_components}, \texttt{min\_dist}, and \texttt{n\_neighbors}.
The \texttt{n\_components} parameter defines the dimensionality of the embedding space. In this study, we set \texttt{n\_components}=2 to project the data onto a two-dimensional plane for visualization.
The \texttt{min\_dist} parameter controls the minimum separation between points in the low-dimensional embedding space. Smaller values result in more compact clusters and place greater emphasis on local structure, while larger values favor the preservation of global structure. 
The \texttt{n\_neighbors} parameter determines the size of the local neighborhood used to approximate the underlying manifold and thus controls the balance between the local and global structure in the embedding space. Smaller values emphasize the local structure, while larger values favor the preservation of global structure.

\begin{figure}[ht!]
\centering
\includegraphics[width=0.48\textwidth]{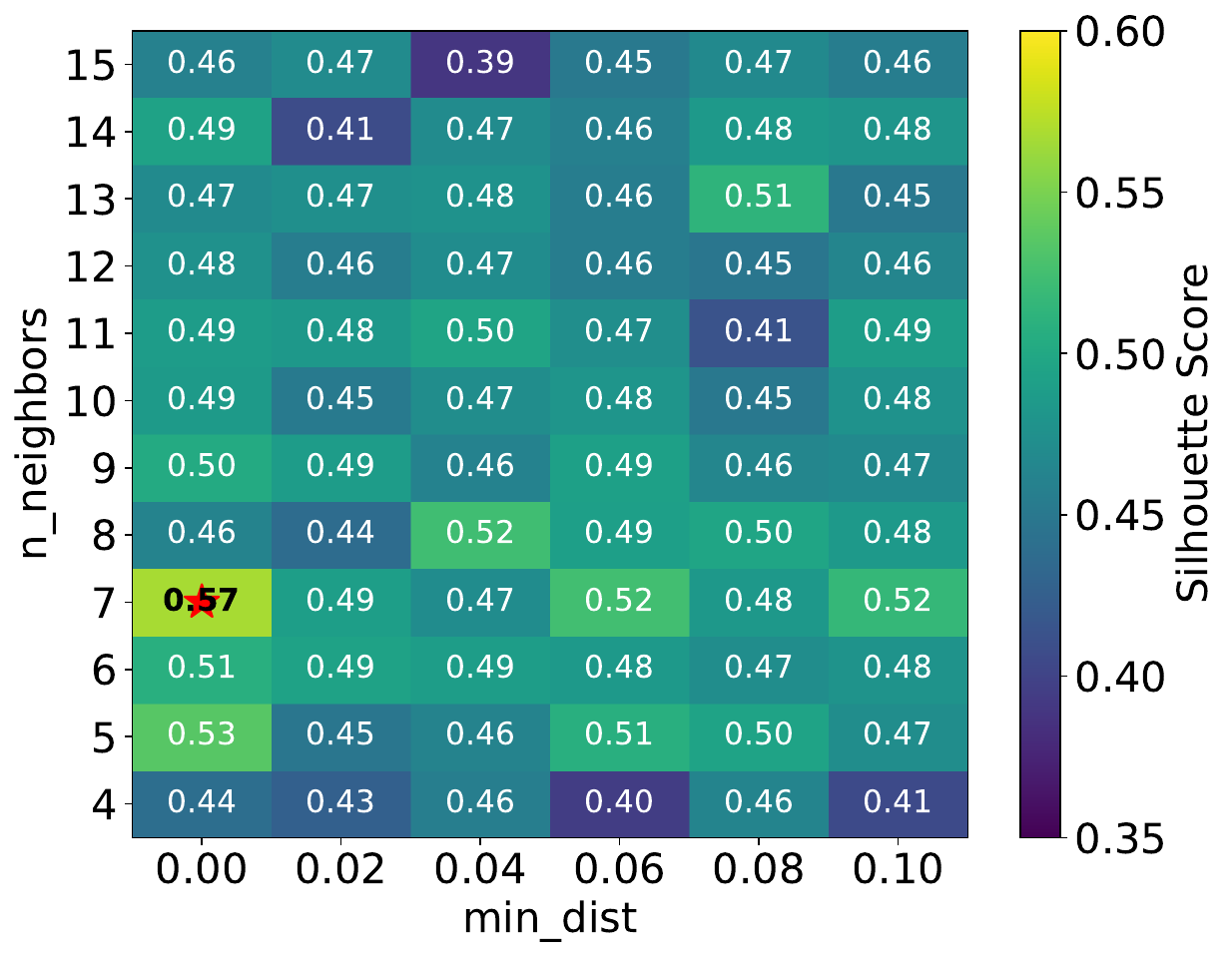}
\includegraphics[width=0.49\textwidth]{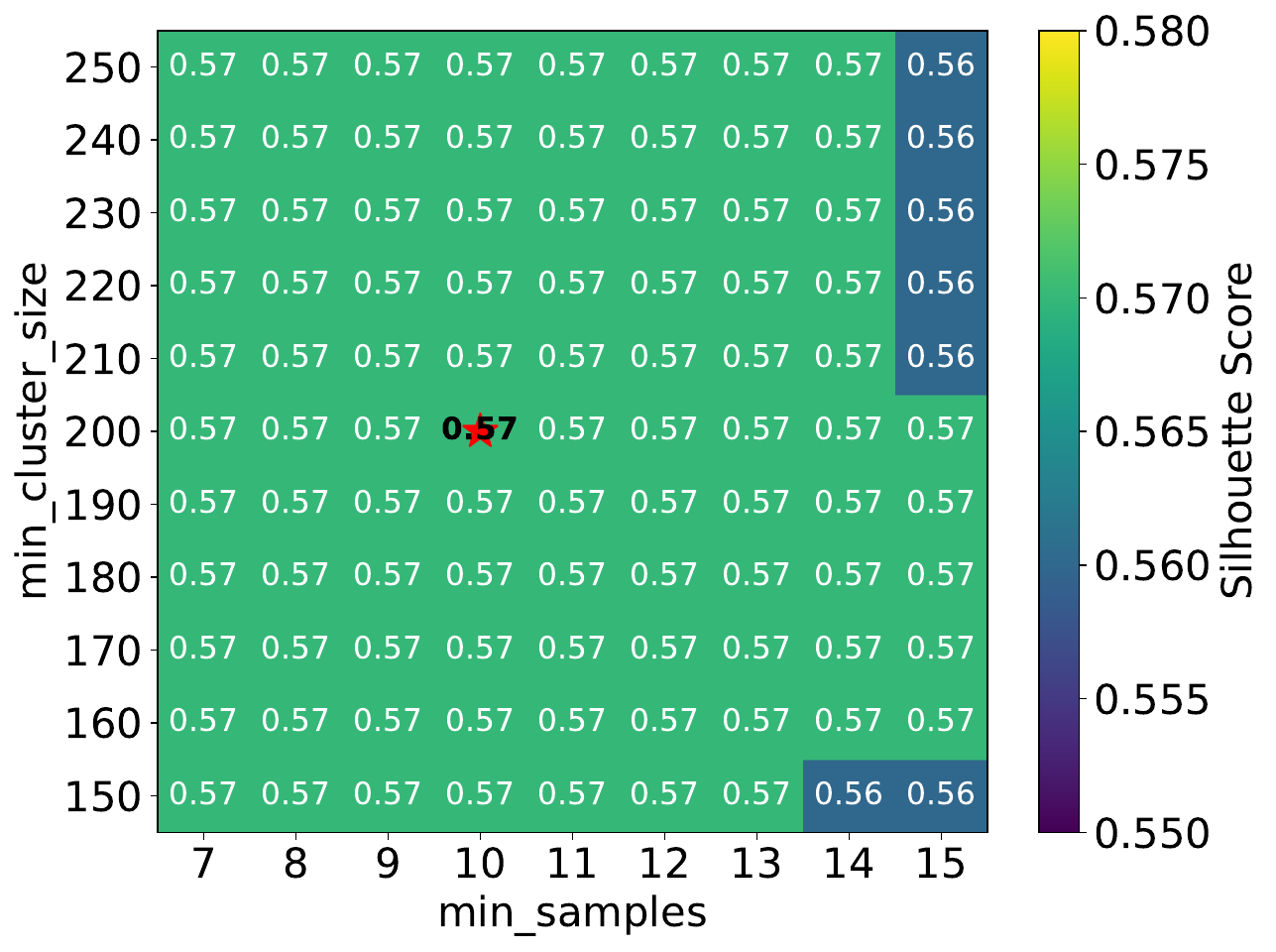}
\caption{The average Silhouette Scores for different hyperparameter configurations. Upper panel: UMAP hyperparameter search with \texttt{min\_cluster\_size}=200 and \texttt{min\_samples}=10. Lower panel: HDBSCAN hyperparameter search with \texttt{n\_neighbors}=7 and \texttt{min\_dist}=0.0. The red star marks the final configuration adopted in this work.}\label{Fig1}
\end{figure}

\subsubsection{Hierarchical Density-Based Spatial Clustering of Applications with Noise}

HDBSCAN is a hierarchical density-based clustering algorithm that identifies clusters based on the density structure of the data. It does not require the number of clusters to be specified in advance and can naturally identify clusters with varying densities and shapes. The algorithm constructs a hierarchical density-based cluster tree and extracts clusters based on their stability across different density levels while assigning isolated or low-density points as noise. This makes HDBSCAN well suited for clustering UMAP embeddings.

HDBSCAN has several important hyperparameters, among which \texttt{min\_cluster\_size} defines the minimum number of data points required to form a cluster. 
Larger values of this parameter typically result in fewer clusters. 
Another important parameter is  \texttt{min\_samples}, which determines the minimum number of neighbors required for a point to be considered a core point. 
Larger values impose stricter density requirements, resulting in more points being classified as noise.

\begin{figure*}[ht!]
\centering
\includegraphics[width=0.45\textwidth]{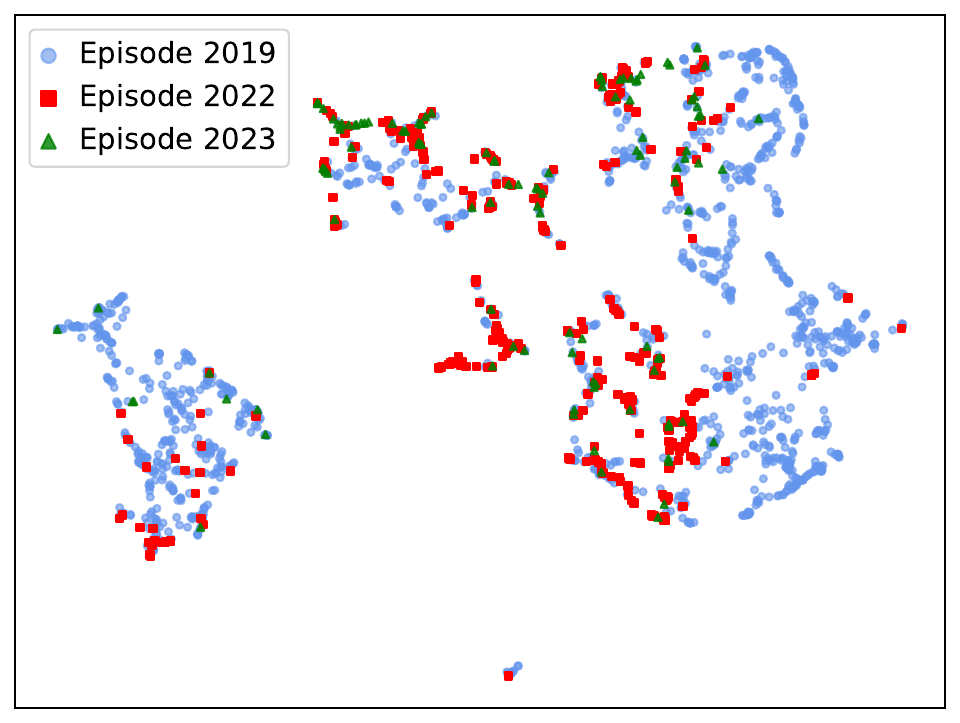}
\hspace{0.03\textwidth} 
\includegraphics[width=0.45\textwidth]{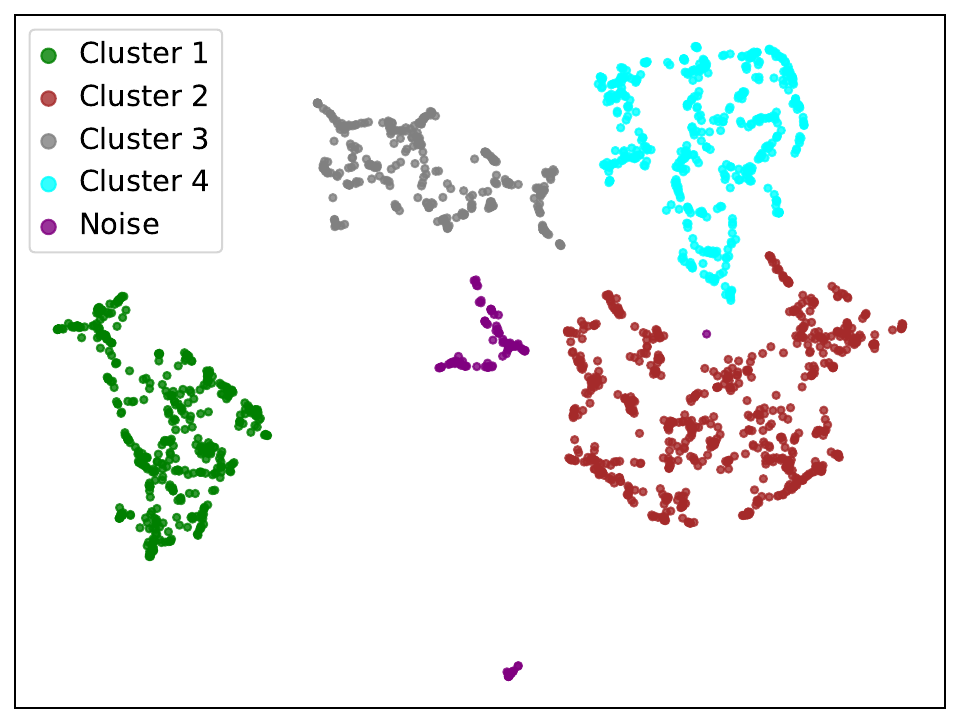}
\caption{The results of dimensionality reduction and clustering. Left panel: the two-dimensional UMAP embedding of bursts from FRB 20121102A. Blue dots, red squares, and green triangles indicate bursts detected from the 2019, 2022, and 2023 active episodes, respectively. Right panel: the HDBSCAN clustering result of the UMAP-embedded data, identifying four distinct clusters with noise.}\label{Fig2}
\end{figure*}

\subsection{Hyperparameter Optimization}\label{sec 2.3}
To determine the optimal UMAP and HDBSCAN hyperparameters, we evaluate the clustering performance of different parameter configurations using the Silhouette Score\footnote{\url{https://scikit-learn.org/stable/modules/generated/sklearn.metrics.silhouette_score.html}} \citep{ROUSSEEUW198753}. 
The Silhouette Score quantifies how well a point fits its cluster and how separated it is from other clusters. For a data point $o_i$, it is defined as
\[s({o_i}) = \frac{{b({o_i}) - a({o_i})}}{{\max \left\{ {a({o_i}),b({o_i})} \right\}}}\]
where $a(o_i)$ is the average distance between $o_i$ and other points within the same cluster, and $b(o_i)$ is the mean distance between $o_i$ and the nearest cluster to which $o_i$ does not belong. The overall Silhouette Score is calculated as the average value of $s(o_i)$ over all data points.
It ranges from $-1$ to $1$, with larger values indicating better-defined clusters. 
We explore the UMAP and HDBSCAN hyperparameters to optimize the clustering performance measured by the Silhouette Score.
The search ranges for \texttt{n\_neighbors}, \texttt{min\_dist}, \texttt{min\_cluster\_size}, and \texttt{min\_samples} are set to $4–15$, $0.0–0.1$, $150–250$, and $7–15$, respectively.
However, a complete four-dimensional grid search requires evaluating a large number of parameter combinations, leading to substantial computational costs and making direct visualization of all configurations impractical. Therefore, we adopt a stepwise optimization strategy. 
Following the hyperparameter choices adopted by \citet{2023MNRAS.524.1668R}, we initially set \texttt{min\_cluster\_size}=200 and \texttt{min\_samples}=10 to ensure that the identified clusters contain a sufficient number of data points while maintaining a balance between cluster detection and noise identification. 
We then search the UMAP hyperparameter space and select the parameter set that yields the highest Silhouette Score. Subsequently, we fix the selected UMAP hyperparameters and further search the HDBSCAN hyperparameter space to identify the optimal configuration.

As shown in the upper panel of Figure \ref{Fig1}, with \texttt{min\_cluster\_size}=200 and \texttt{min\_samples}=10, the combination of \texttt{n\_neighbors}=7 and \texttt{min\_dist}=0.0 yields the highest Silhouette Score. 
We therefore adopt these UMAP hyperparameters for subsequent HDBSCAN optimization.
As shown in the lower panel of Figure \ref{Fig1}, the maximum Silhouette Score is maintained within the parameter range of \texttt{min\_cluster\_size}=150–250 and \texttt{min\_samples}=7–13, indicating that the clustering performance is stable in this region. Furthermore, we select four HDBSCAN hyperparameter combinations from this region to evaluate the robustness of the selected UMAP hyperparameters. The results presented in Appendix \ref{app1} (Figure \ref{Fig11}) demonstrate that the UMAP configuration with \texttt{n\_neighbors}=7 and \texttt{min\_dist}=0.0 provides robust clustering performance across different HDBSCAN hyperparameter combinations. Therefore, we adopt the final hyperparameter configuration of \texttt{n\_neighbors}=7, \texttt{min\_dist}=0.0, \texttt{min\_cluster\_size}=200, and \texttt{min\_samples}=10.

All analyses presented in this work were performed using Python \texttt{3.10.7}. The main libraries included \texttt{scikit-learn 1.7.1} \citep{1953048.2078195} for feature preprocessing and standardization, as well as calculation of the Silhouette Score; \texttt{umap-learn 0.5.11} \citep{2018arXiv180203426M} for nonlinear dimensionality reduction; and \texttt{hdbscan 0.8.41} \citep{2017JOSS....2..205M} for density-based clustering. Other libraries, including \texttt{numpy 2.2.5}, \texttt{pandas 2.2.3}, and \texttt{matplotlib 3.10.1}, were used for numerical computation, data processing, and data visualization, respectively.

\begin{table*}[htbp]
\centering
\small
\setlength{\tabcolsep}{2pt}  
\caption{The mean values of the eight burst parameters for the four clusters and Noise.}
\label{Tab1}

\begin{tabular}{lcccccccccccc}
\toprule
Cluster 
& Width 
& Bandwidth 
& Central frequency 
& Bandwidth ratio 
& Peak flux 
& Fluence 
& Waiting time 
& $\log_{10} \rm {Energy}$ \\

& (ms) 
& (MHz) 
& (MHz) 
& 
& (mJy) 
& (mJy\,ms) 
& (s) 
& (erg) \\
\midrule
Noise & 8.56 $\pm$ 4.87 & 188 $\pm$ 56 & 1377 $\pm$ 46 & 0.14 $\pm$ 0.04 & 17.63 $\pm$ 10.69 & 132 $\pm$ 62 & 305 $\pm$ 735 & 38.03 $\pm$ 0.20 \\
1 & 5.07 $\pm$ 5.43 & 233 $\pm$ 126 & 1310 $\pm$ 89 & 0.18 $\pm$ 0.10 & 155.16 $\pm$ 217.16 & 606 $\pm$ 785 & 100 $\pm$ 151 & 38.64 $\pm$ 0.34 \\
2 & 4.46 $\pm$ 2.35 & 146 $\pm$ 60 & 1377 $\pm$ 44 & 0.11 $\pm$ 0.04 & 15.02 $\pm$ 8.15 & 58 $\pm$ 24 & 76 $\pm$ 105 & 37.72 $\pm$ 0.17 \\
3 & 5.43 $\pm$ 3.04 & 356 $\pm$ 75 & 1279 $\pm$ 47 & 0.28 $\pm$ 0.06 & 20.70 $\pm$ 14.42 & 99 $\pm$ 73 & 51 $\pm$ 79 & 37.94 $\pm$ 0.27 \\
4 & 5.64 $\pm$ 2.57 & 163 $\pm$ 58 & 1181 $\pm$ 75 & 0.14 $\pm$ 0.05 & 12.95 $\pm$ 6.32 & 69 $\pm$ 42 & 82 $\pm$ 105 & 37.79 $\pm$ 0.20 \\

\bottomrule
\end{tabular}
\end{table*}

\begin{figure*}[ht!]
\centering
\includegraphics[width=\textwidth,keepaspectratio]{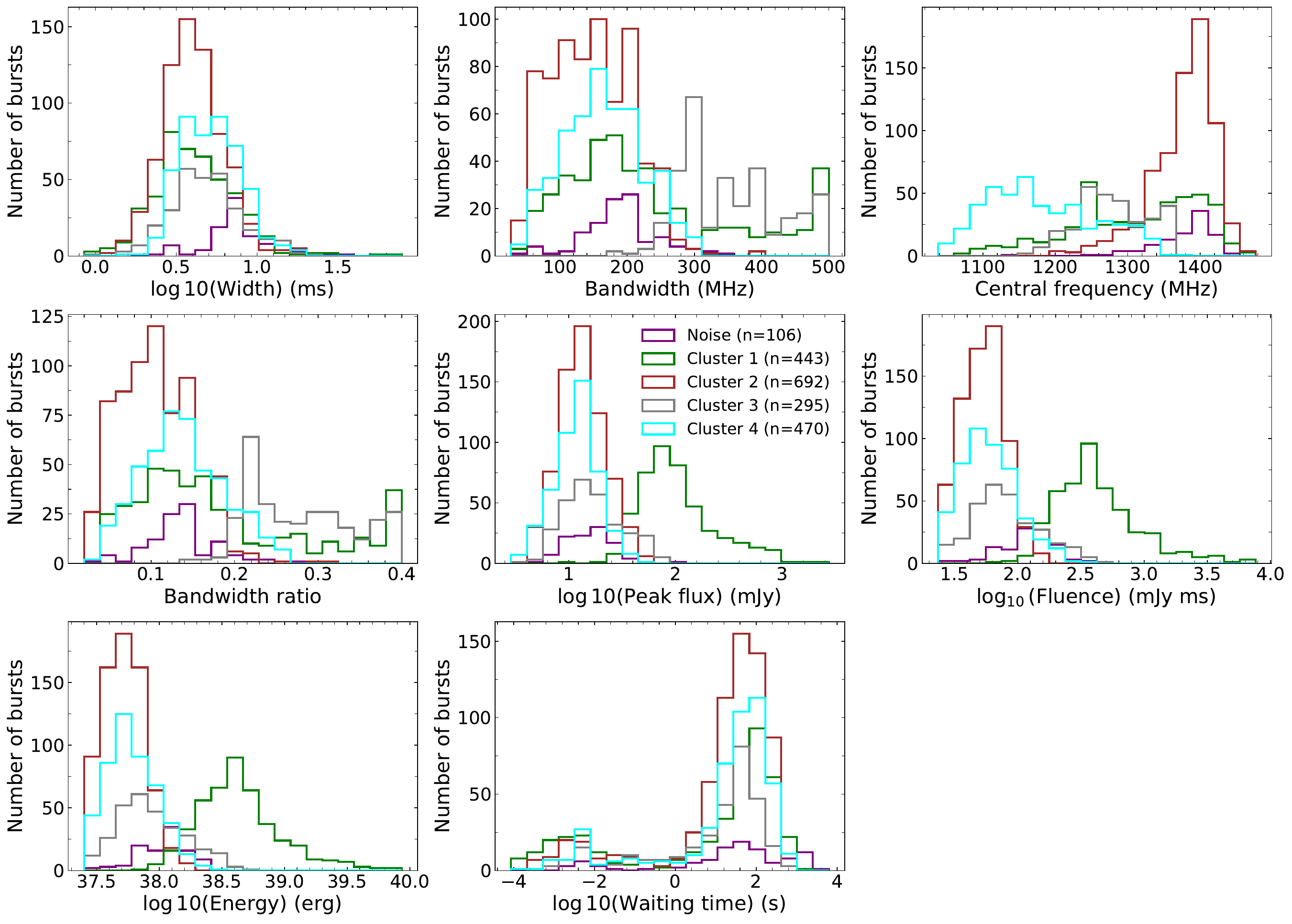}
\caption{The distributions of the eight burst parameters for different clusters. The green, orange, gray, cyan and purple histograms represent Cluster 1, Cluster 2, Cluster 3, Cluster 4, and Noise, respectively.}\label{Fig3}
\end{figure*}

\section{Results}\label{sec3}
The two-dimensional UMAP embedding of bursts from FRB 20121102A across three distinct active episodes is shown in the left panel of Figure \ref{Fig2}. 
The blue dots represent bursts from the 2019 active episode, while red squares and green triangles denote bursts from the 2022 and 2023 active episodes, respectively.
It is evident that the bursts from the 2019 active episode occupy a broader region of the embedding space, whereas those from the 2022 and 2023 active episodes are mainly concentrated in the upper and central regions. This suggests that bursts from different active episodes exhibit distinct distributions in the embedding space, indicating variations in their underlying multidimensional burst properties. We then apply HDBSCAN to cluster the two-dimensional UMAP embedding, and the results are shown in the right panel of Figure \ref{Fig2}. HDBSCAN identifies four distinct clusters with noise under the adopted hyperparameter configuration. The four clusters exhibit clearly separated distributions in the UMAP embedding space, while the noise points are not assigned to any of these clusters.
For visualization, the four clusters and the noise points are assigned distinct colors and labeled as Cluster 1, Cluster 2, Cluster 3, Cluster 4, and Noise, respectively. 

The clustering results in the embedding space reveal the diversity of FRB burst properties. We further examine the distributions of the eight parameters used as input features for UMAP to investigate the characteristics of each cluster. The mean values of these parameters for the four clusters and the noise points are summarized in Table \ref{Tab1}, and their distributions are presented in Figure \ref{Fig3}. We find that bursts in Cluster 1 exhibit substantially different distributions of energy, peak flux, and fluence compared with those in the other three clusters. The mean values of these parameters are approximately one order of magnitude higher than those of the other clusters. 
In addition, the distributions of central frequency, bandwidth, and bandwidth ratio for Cluster 1 bursts display bimodal-like features that are not observed in the other clusters.
Although bursts in Clusters 2 and 4 have similar bandwidth ratios, with mean values of $\sim$ 0.1, they exhibit distinct spectral characteristics. Bursts in Cluster 2 are distributed toward higher central frequencies, whereas those in Cluster 4 are preferentially located at lower central frequencies, despite having comparable bandwidths. In contrast, bursts in Cluster 3 exhibit relatively broad bandwidths, intermediate central frequencies, and larger bandwidth ratios. The pronounced differences in central frequency, bandwidth, and bandwidth ratio among clusters indicate that spectral properties may be an important characteristic for distinguishing different burst clusters. The width distribution of the noise points differs markedly from those of the four clusters, whose width distributions are generally similar to one another. The central frequency and bandwidth distributions of the noise points, however, are broadly similar to those of Cluster 2.
Additionally, the waiting time distributions are similar across the four clusters, all of which exhibit bimodal distributions.

\begin{figure}[ht!]
\centering
\includegraphics[width=0.45\textwidth]{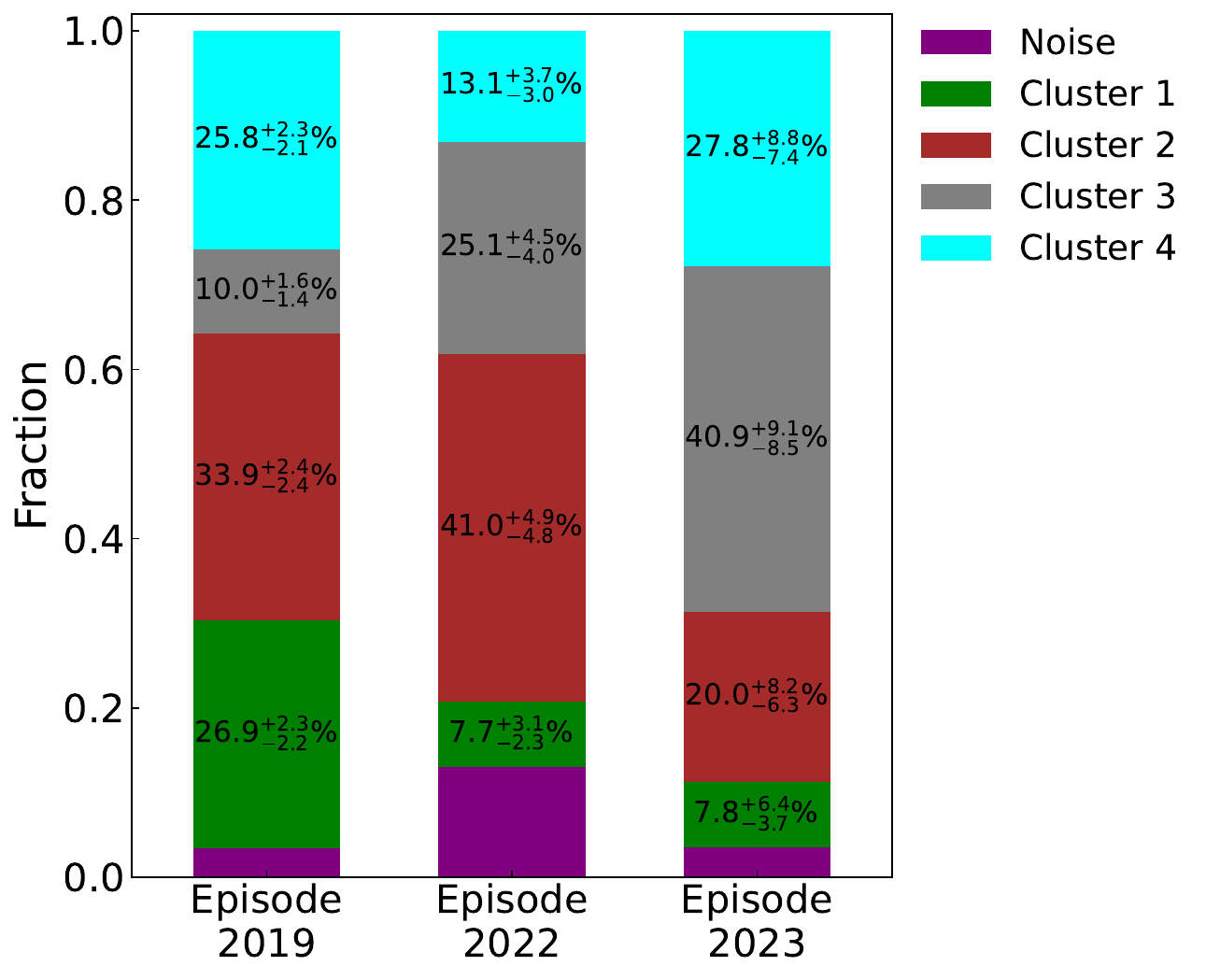}
\caption{The fractions of bursts in each cluster across three active episodes. The uncertainties represent the 95\% Wilson score confidence intervals. }\label{Fig4}
\end{figure}

\begin{figure*}[ht!]
\centering
\includegraphics[width=0.45\textwidth]{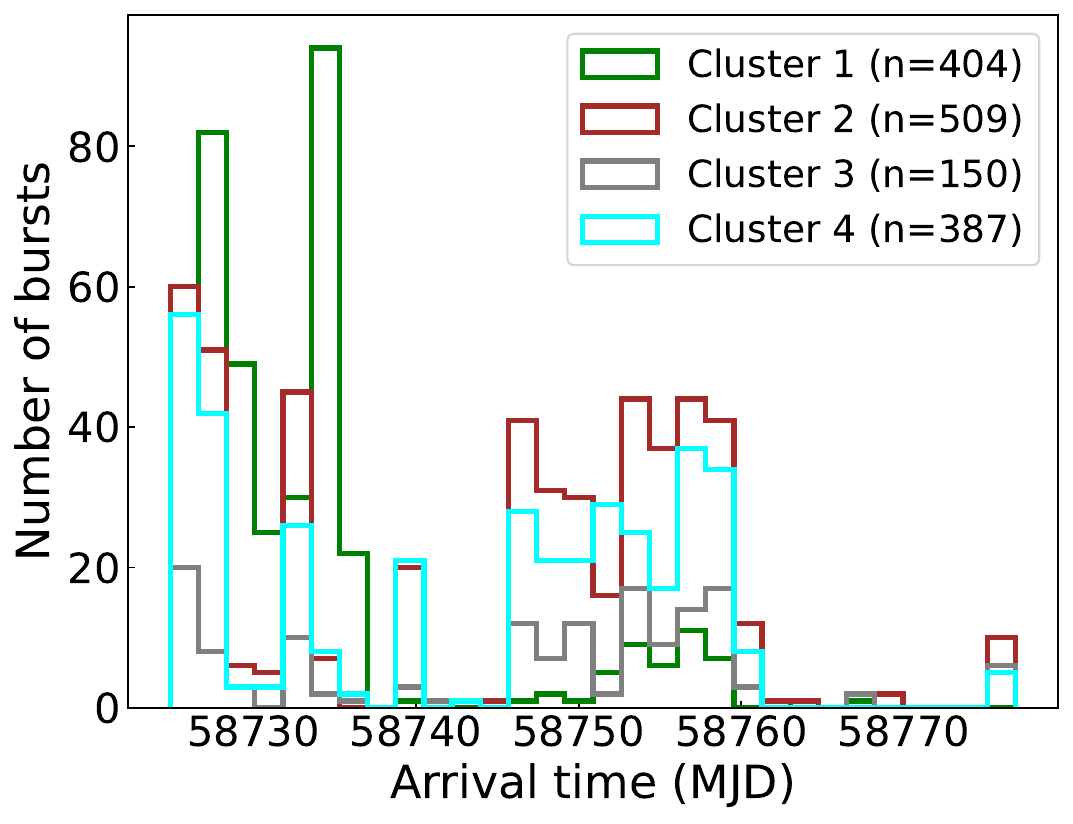}
\hspace{0.03\textwidth} 
\includegraphics[width=0.45\textwidth]{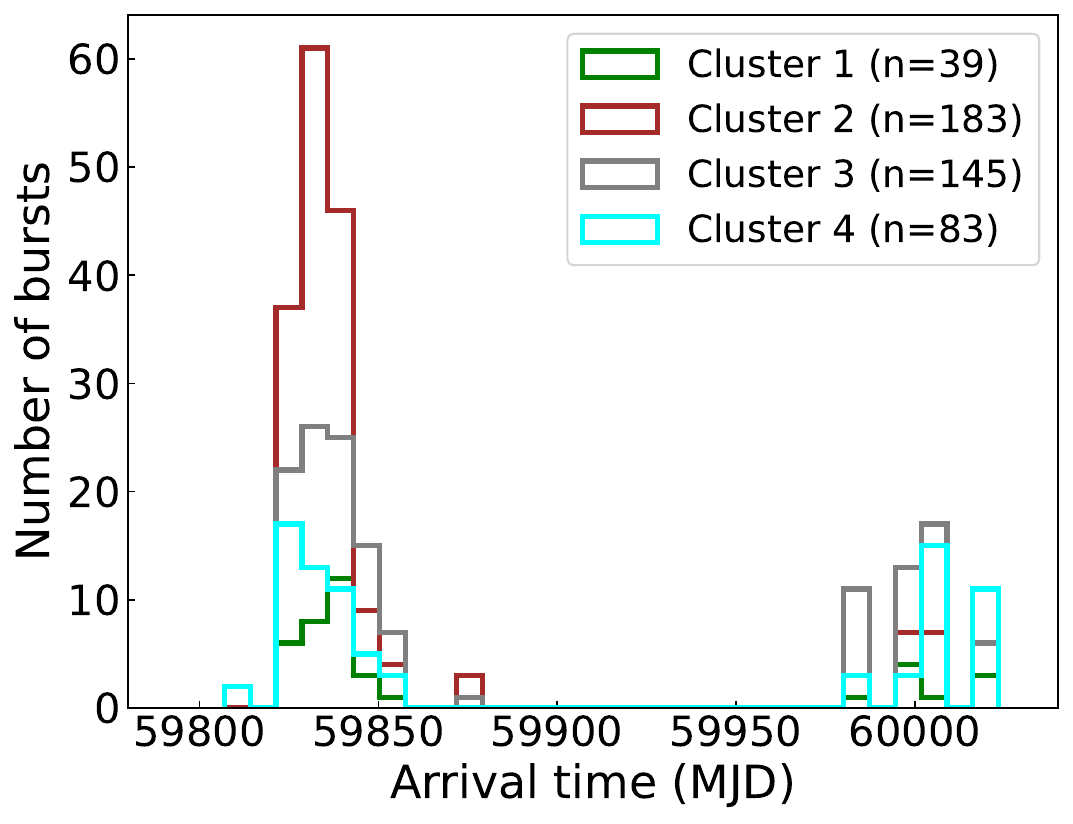}
\caption{ The arrival time distributions of bursts in the four clusters during the 2019 active episode (left panel) and the 2022 and 2023 active episodes (right panel).}\label{Fig5}
\end{figure*}

To further investigate whether the burst populations vary across active episodes, we examine the fraction of bursts assigned to the different clusters in each active episode.
Table \ref{Tab2} summarizes the number of bursts in each cluster during each active episode, while the corresponding fractions are shown in Figure \ref{Fig4}. The statistical uncertainties in these fractions are quantified using 95\% Wilson confidence intervals for binomial proportions. Cluster 1 accounts for $26.9^{+2.3}_{-2.2}\%$ of the bursts in the 2019 active episode, compared with $7.7^{+3.1}_{-2.3}\%$ and $7.8^{+6.4}_{-3.7}\%$ in the 2022 and 2023 active episodes, respectively.
In addition, Cluster 2 contains the largest fraction of bursts in the 2019 and 2022 active episodes, whereas Cluster 3 contains the largest fraction in the 2023 active episode.   We also examine whether bursts in different clusters exhibit temporal dependence within each active episode. The left and right panels of Figure \ref{Fig5} present the arrival time distributions of bursts in the four clusters during the 2019 active episode and the 2022 and 2023 active episodes, respectively. The results show that, within each active episode, bursts from different clusters do not show apparent time dependence.

\begin{table}[htbp]
\centering
\small
\caption{The number of bursts in each cluster and Noise across three active episodes.}
\label{Tab2}
\setlength{\tabcolsep}{7pt}
\renewcommand{\arraystretch}{1.1}
\begin{tabular}{cccc}
\toprule
Cluster & Episode 2019 & Episode 2022 & Episode 2023 \\
\midrule
Noise & 51  & 51  & 4 \\
1 & 404  & 30  & 9  \\
2 & 509 & 160 & 23 \\
3 & 150 & 98 & 47 \\
4 & 387 & 51  & 32 \\
\bottomrule
\end{tabular}
\end{table}

\section{Discussions}\label{sec4}
\subsection{Reliability and stability of the clustering structure}
\begin{figure}[ht!]
\centering
\includegraphics[width=0.45\textwidth]{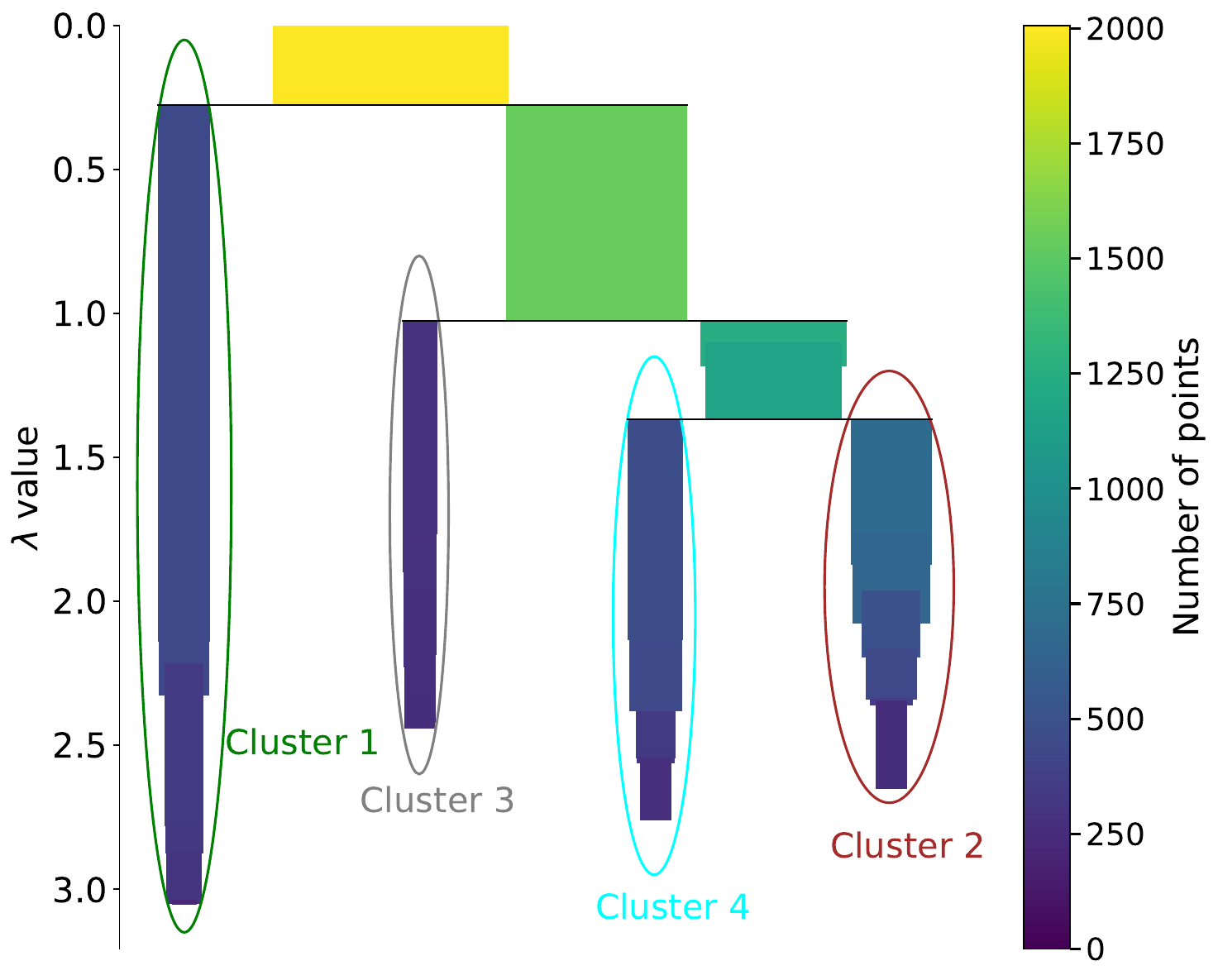}
\caption{The HDBSCAN condensed tree obtained with \texttt{n\_neighbors}=7, \texttt{min\_dist}=0.0, \texttt{min\_cluster\_size}=200, and \texttt{min\_samples}=10. Leaf nodes enclosed by ellipses of different colors represent the final clusters identified by HDBSCAN.
}\label{Fig6}
\end{figure}

Our analysis adopts the UMAP and HDBSCAN hyperparameter configuration that yields the highest Silhouette Score. To examine the hierarchical structure and persistence of the identified clusters, we present the corresponding HDBSCAN condensed tree for the adopted hyperparameter configuration in Figure \ref{Fig6}.
The condensed tree shows the hierarchical splitting of the data as the density level $\lambda$ increases, where $\lambda=1/d$ and $d$ is the distance between a point and cluster core.
Cluster 1 separates from the main structure at $\lambda\sim0.3$ and persists to $\lambda\sim3.0$, while the remaining data subsequently separate into Cluster 3 at $\lambda\sim1.0$ and further split into Clusters 4 and 2. The four identified clusters can be traced across distinct ranges of density levels in the condensed tree, supporting the persistence of the identified clustering structure.

\begin{figure*}[ht!]
\centering
\includegraphics[width=0.48\textwidth]{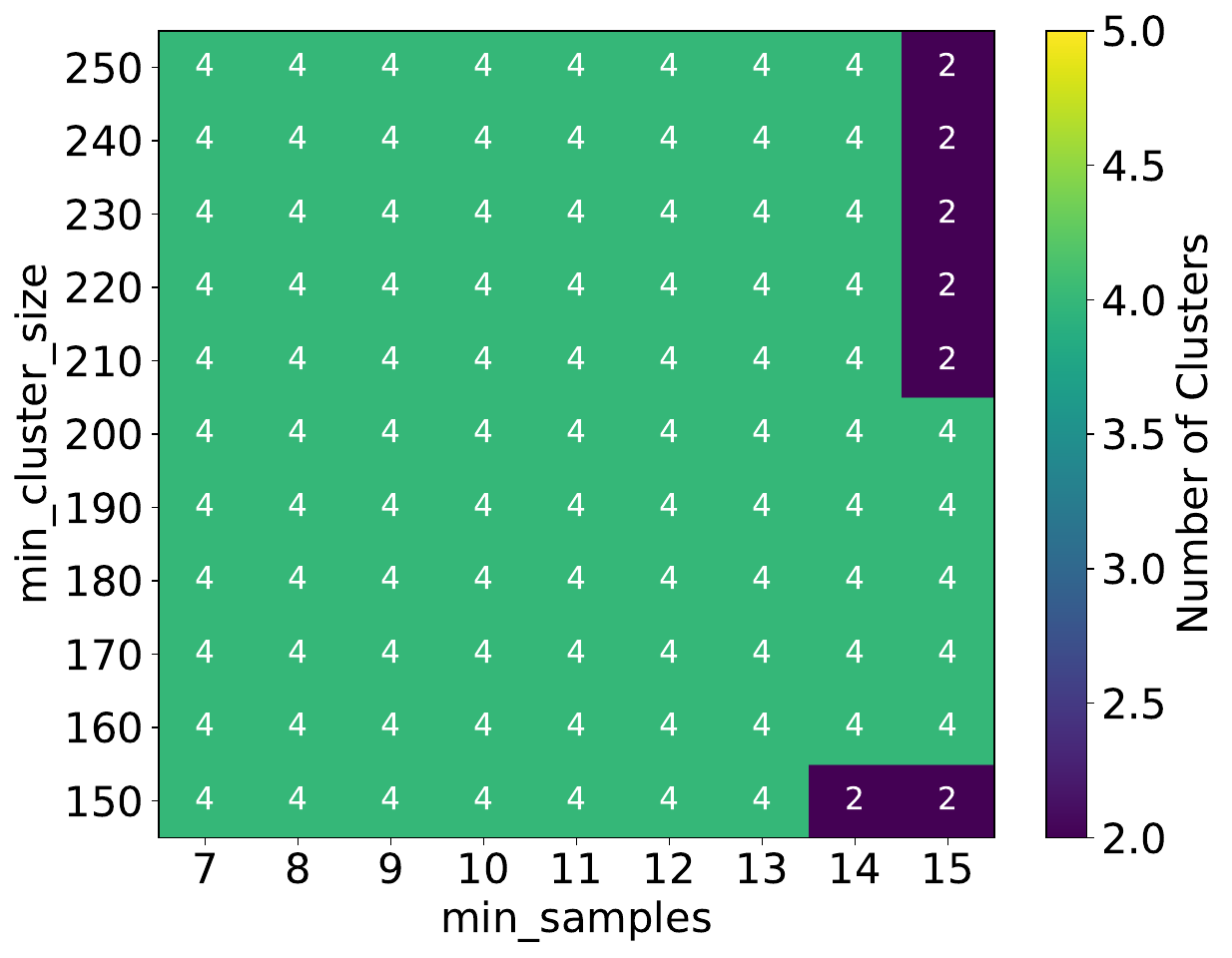}
\includegraphics[width=0.49\textwidth]{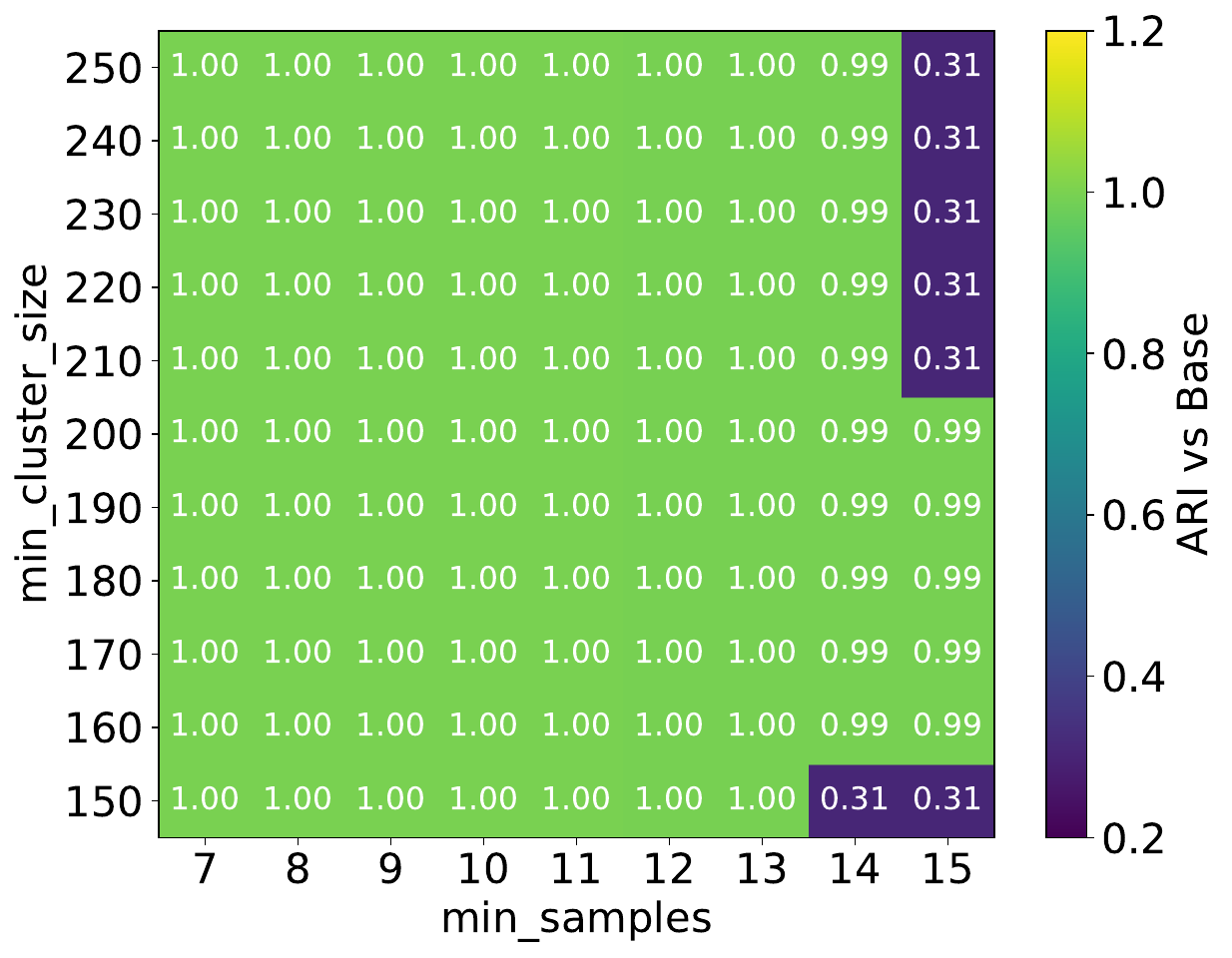}
\caption{The number of clusters identified (left) for different HDBSCAN hyperparameter combinations, with the UMAP hyperparameters \texttt{n\_neighbors}=7 and \texttt{min\_dist}=0.0 fixed, and the corresponding Adjusted Rand Index (ARI) (right) relative to the final adopted configuration (\texttt{n\_neighbors}=7, \texttt{min\_dist}=0.0, \texttt{min\_cluster\_size}=200, and \texttt{min\_samples}=10).}\label{Fig7}
\end{figure*}

\begin{figure*}[ht!]
\centering
\includegraphics[width=\textwidth,keepaspectratio]{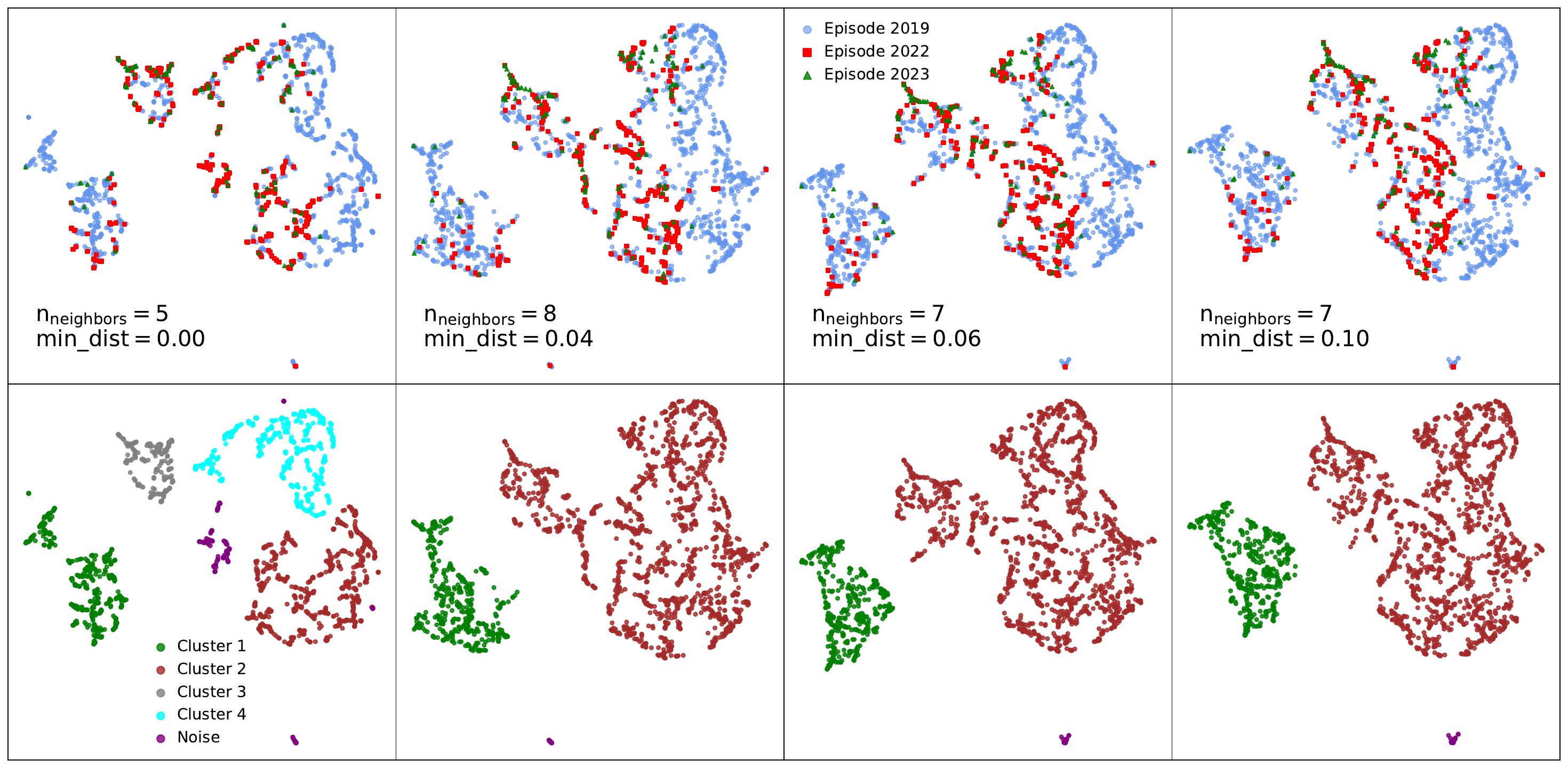}
\caption{The UMAP embeddings (upper panels) and HDBSCAN clustering results (lower panels) for different UMAP hyperparameter combinations, with the HDBSCAN hyperparameters \texttt{min\_cluster\_size}=200 and \texttt{min\_samples}=10 fixed.}\label{Fig8}
\end{figure*}

We further investigate the robustness of the clustering to changes in the HDBSCAN hyperparameters.  The left panel of Figure \ref{Fig7} shows the number of clusters identified for \texttt{min\_cluster\_size}=150–250 and \texttt{min\_samples}=7–15, with the UMAP hyperparameters \texttt{n\_neighbors}=7 and \texttt{min\_dist}=0.0 fixed. The right panel of Figure \ref{Fig7} presents the Adjusted Rand Index\footnote{\url{https://scikit-learn.org/stable/modules/generated/sklearn.metrics.adjusted_rand_score.html}} (ARI; \citealt{steinley2004properties}) between the clustering obtained for each HDBSCAN hyperparameter combination and that obtained with the adopted configuration. The ARI measures the similarity between two clusterings, with values closer to 1 indicating greater consistency in their cluster assignments. For the hyperparameter range of \texttt{min\_cluster\_size}=150-250 and \texttt{min\_samples}=7-13, where the maximum Silhouette Score is maintained in Section \ref{sec 2.3}, all hyperparameter combinations consistently identify four clusters, with ARI values close to 1.
These results indicate that the clustering structure remains stable over the tested range of HDBSCAN hyperparameters, supporting the robustness of the identified four clusters.
We also assess the dependence of the clustering results on the UMAP hyperparameters, as nonlinear dimensionality reduction techniques are known to be sensitive to hyperparameter choices and initialization and can affect the representation of the global data structure \citep{kobak2021initialization,JMLR:v22:20-1061}. 
As shown in Appendix \ref{app1} and the upper panel of Figure \ref{Fig1}, the UMAP configuration with \texttt{n\_neighbors}=7 and \texttt{min\_dist}=0.0 consistently yields the highest Silhouette Score across different HDBSCAN hyperparameter combinations. We fix the HDBSCAN hyperparameters at \texttt{min\_cluster\_size}=200 and \texttt{min\_samples}=10 and select the four UMAP hyperparameter combinations with the highest Silhouette Scores among the remaining configurations, namely (\texttt{n\_neighbors}, \texttt{min\_dist})=(5,0.0), (8,0.04), (7,0.06), and (7,0.1). 
The upper and lower panels of Figure \ref{Fig8} show the two-dimensional UMAP embeddings and the corresponding HDBSCAN clustering results, respectively. The embeddings show similar overall structures, with 2019 bursts occupying a broader region, and 2022 and 2023 bursts mainly concentrated in the upper and central regions. In contrast, the clustering structure is more sensitive to the UMAP hyperparameters. Except for the configuration with \texttt{n\_neighbors}=5 and \texttt{min\_dist}=0.0, which divides the bursts into four clusters with noise, the other three hyperparameter combinations identify two clusters with noise. In Appendix \ref{app2}, we compare the feature distributions of the clusters and their fractions across three active episodes for the \texttt{n\_neighbors}=5 and \texttt{min\_dist}=0.0 configuration, finding that they are consistent with those obtained for the adopted configuration.

\subsection{Feature importance analysis} 
After dimensionality reduction and clustering, the relationship between the identified clusters and the input features remains unclear.
To investigate the burst properties associated with these clusters, we perform a mutual information (MI) analysis between the input features and the UMAP embeddings, as well as the HDBSCAN clustering results.
MI measures the dependence between two variables,  with higher values indicating stronger dependencies \citep{10.1371/journal.pone.0087357}. 
For UMAP embedding, MI scores are calculated between each input feature and the two-dimensional embedding coordinates within each cluster using \texttt{mutual\_info\_regression}\footnote{\url{https://scikit-learn.org/stable/modules/generated/sklearn.feature_selection.mutual_info_regression.html}}. 
For HDBSCAN results, each cluster is considered independently against the remaining clusters, and \texttt{mutual\_info\_classif}\footnote{\url{https://scikit-learn.org/stable/modules/generated/sklearn.feature_selection.mutual_info_classif.html}} is used to quantify the association between the input features and cluster membership. 
The resulting MI scores for the UMAP embedding coordinates and HDBSCAN cluster memberships are stacked and visualized using bar plots with different colors.

\begin{figure}[ht!]
\centering
\includegraphics[width=0.45\textwidth]{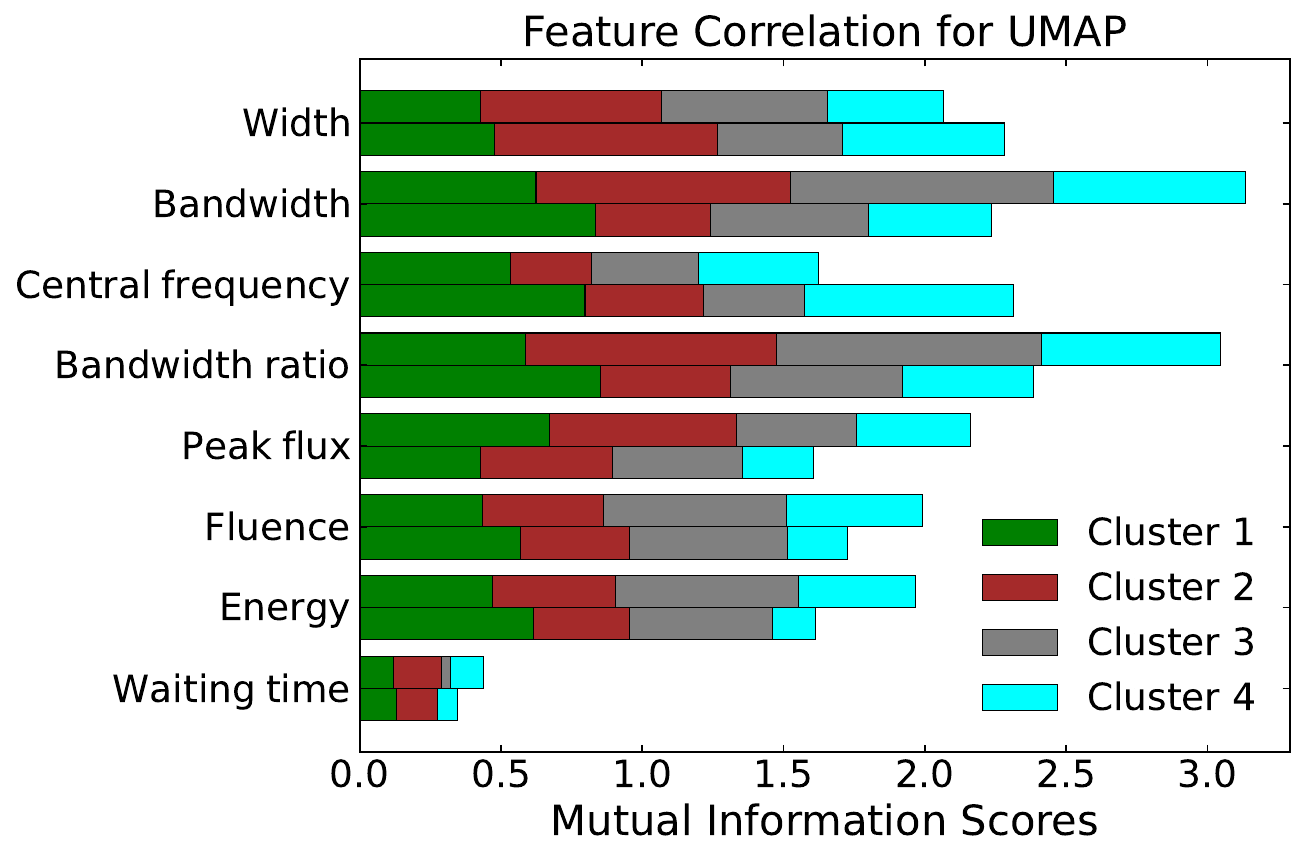}
\includegraphics[width=0.45\textwidth]{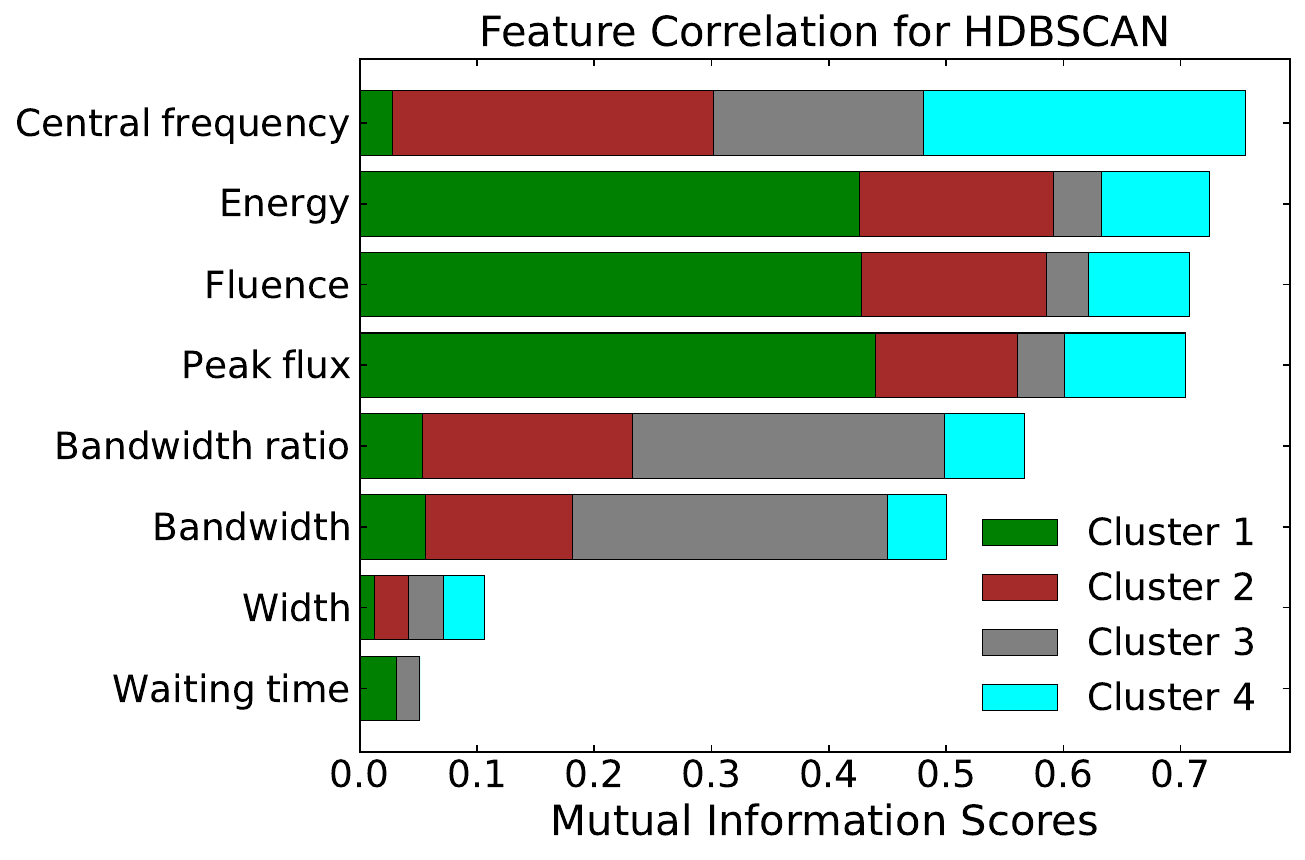}
\caption{Mutual information scores between the input features and the UMAP embedding coordinates (upper panel) and HDBSCAN cluster memberships (lower panel). For UMAP, the top and bottom bars represent the horizontal and vertical embedding coordinates, respectively.}\label{Fig9}
\end{figure}

The upper panel of Figure \ref{Fig9}  presents the MI scores between the input features and the UMAP embedding coordinates for each cluster. All four clusters exhibit  high MI scores for width, central frequency, bandwidth, bandwidth ratio, energy, fluence, and peak flux, whereas waiting time shows consistently low MI values across all clusters.  The lower panel of Figure \ref{Fig9} shows the MI scores between the input features and individual HDBSCAN cluster memberships. Cluster 1 exhibits the highest MI values for energy, fluence, and peak flux, whereas the other clusters show higher MI values primarily for central frequency, bandwidth, and bandwidth ratio. In contrast, width and waiting time show consistently low MI values across all cluster memberships. By combining the MI results from UMAP and HDBSCAN, we find that, except for waiting time, all input features show relatively strong associations with the distribution of bursts in the UMAP embedding, whereas their associations with the HDBSCAN cluster assignments exhibit clear variations across different features and clusters. Specifically, spectral properties, characterized by central frequency, bandwidth, and bandwidth ratio, play a dominant role in distinguishing different burst clusters of FRB 20121102A, while energy, peak flux, and fluence are mainly associated with Cluster 1. These results are consistent with the observed cluster properties: apart from the distinct energetic properties of Cluster 1, the differences among burst clusters are primarily reflected in their distributions of central frequency, bandwidth, and bandwidth ratio.

\subsection{Comparison with other results}
\begin{figure}[ht!]
\centering
\includegraphics[width=0.45\textwidth]{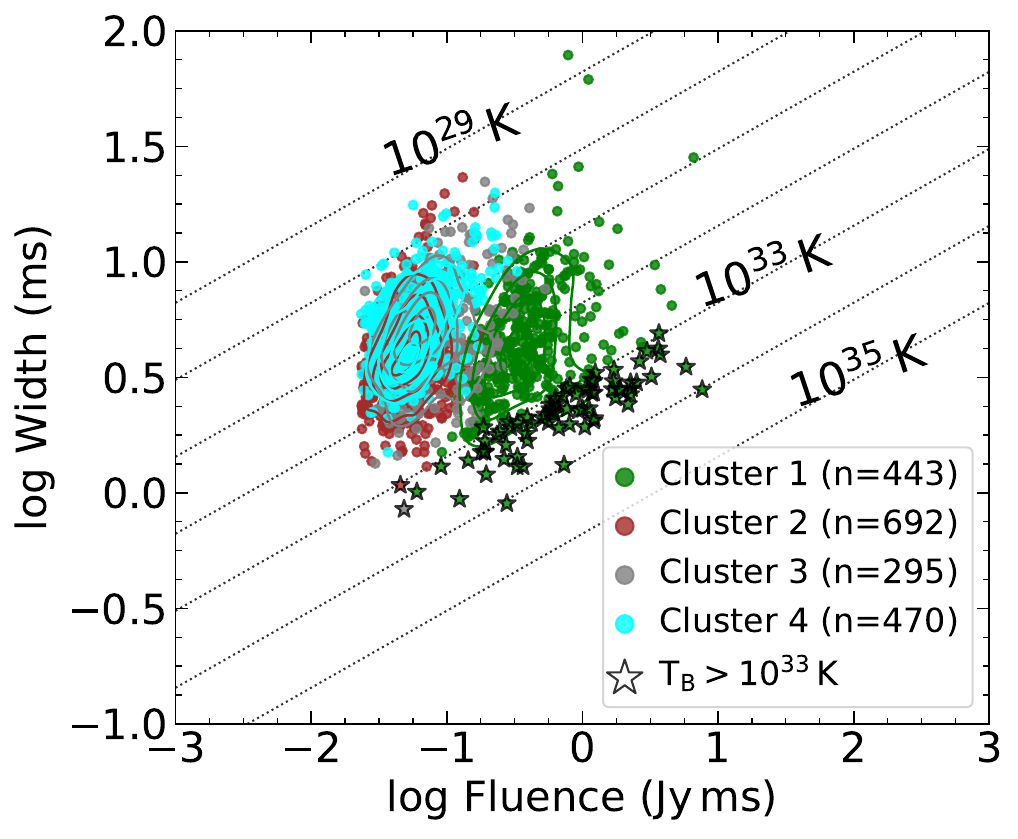}
\caption{Scatter plot of width versus fluence for the bursts from four different clusters. The stars mark bursts with $T_{\rm {B}}>10^{33}$ K, corresponding to the ``classical'' bursts defined by \cite{2022A&A...657L...7X}. The dotted lines show the width--fluence relations for different fixed $T_{\rm B}$, calculated using Equation (1) of \cite{2022A&A...657L...7X} and adopting the median central frequency of all bursts from three active episodes.}\label{Fig10}
\end{figure}

Previous studies have suggested that the bursts from FRB 20121102A may contain multiple distinct subpopulations. Based on the critical brightness temperature ($T_{\mathrm{B,cri}}\approx10^{33}$ K), \cite{2022A&A...657L...7X} classified the 1652 bursts reported by \cite{2021Natur.598..267L} into ``classical'' and ``atypical'' FRBs. Bursts with $T_{\rm {B}}\geq10^{33}$ K were classified as ``classical'' FRBs, whereas those with $T_{\rm{B}}<10^{33}$ K were classified as ``atypical'' FRBs. They further found that the ``classical'' bursts follow a tight width--fluence relation. To compare their classification with our clustering results, we project the four clusters identified in this work onto the width--fluence parameter space. 
As shown in Figure \ref{Fig10}, Cluster 1 contains the majority of the ``classical'' bursts identified by \cite{2022A&A...657L...7X}.
In addition, Cluster 1 occupies a distinct region in the width--fluence parameter space compared with the other clusters.
For comparison, \cite{2023MNRAS.524.1668R} applied a UMAP–HDBSCAN clustering framework using six observational parameters to the FAST burst sample of FRB 20121102A reported by \cite{2021Natur.598..267L} and identified three clusters with noise. 
Our results are broadly consistent with the findings of \cite{2023MNRAS.524.1668R} that energy, fluence, peak flux, and bandwidth are important for distinguishing different burst clusters.
However, by including central frequency and adopting an expanded feature space, we further identify four clusters with clearer spectral differences. 
In particular, Cluster 1 identified by \cite{2023MNRAS.524.1668R}, characterized by low bandwidth and low energy, exhibits properties broadly similar to our Clusters 2 and 4, suggesting that the absence of central frequency may reduce the separability of different spectral populations.
The importance of spectral parameters is further supported by \cite{2024PASA...41...90L}. 
Using 977 bursts detected by the Arecibo telescope, they applied an unsupervised learning approach with seven observational parameters and found that different clusters are mainly distinguished by spectral properties, particularly central frequency and bandwidth.  
This result is consistent with our analysis, indicating that spectral properties plays a crucial role in revealing the complex substructures of repeating FRBs.

\subsection{Physical Interpretation of Clusters}\label{sec4.3}
Our analysis shows that the bursts in Cluster 1 exhibit significant differences from those in the other clusters. This cluster contains almost all high-energy bursts and shows bimodal-like distributions in both central frequency and bandwidth. 
The high-energy component of the bimodal energy distribution reported by \cite{2021Natur.598..267L} is almost entirely associated with Cluster 1, and the majority of the ``classical'' bursts identified by \cite{2022A&A...657L...7X} are also distributed in this cluster. 
In addition, Cluster 1 occupies a distinct region in the width--fluence parameter space compared with the other clusters. These results suggest that Cluster 1 may represent a physically distinct burst population,
potentially associated with different emission conditions or radiation mechanisms.
The radiation mechanism of FRBs remains unknown. However, the high brightness temperature of FRBs implies coherent emission \citep{2020Natur.587...45Z,2021Univ....7...56L,2021SCPMA..6449501X}. 
Proposed mechanisms include synchrotron maser radiation from relativistic shocks (e.g., \citealt{2014MNRAS.442L...9L,2017ApJ...843L..26B,2020ApJ...896..142B,2019MNRAS.485.4091M,2017ApJ...842...34W,2025A&A...695A.100L}) and coherent curvature radiation or inverse Compton scattering by particle bunches in the magnetosphere (e.g., \citealt{2014PhRvD..89j3009K,2017MNRAS.468.2726K,2018ApJ...868...31Y,2022ApJ...925...53Z,2024ApJ...972..124Q}).
The high-energy bursts may indicate different physical conditions under different radiation scenarios. In the synchrotron maser model, they could be associated with more energetic magnetar flares or a more efficient conversion of shock energy into coherent radio emission \citep{2014MNRAS.442L...9L,2017ApJ...842...34W,2019MNRAS.485.4091M}. In magnetospheric models, they may instead arise from more efficient particle acceleration or more extreme plasma conditions \citep{2017MNRAS.468.2726K,2022ApJ...925...53Z}.
In contrast, the other clusters are primarily distinguished by spectral properties, including central frequency, bandwidth, and bandwidth ratio. 
These differences may be related to the geometry or local physical conditions of the emission regions. 
In curvature radiation models, the characteristic frequency follows $\nu_{\rm c}\propto\gamma^3/\rho$, where $\gamma$ and $\rho$ represent the particle Lorentz factor and curvature radius of magnetic field lines, respectively \citep{2017MNRAS.468.2726K}. 
Different emission regions can naturally lead to spectral differences, with higher-frequency bursts (e.g., Cluster 2) potentially originating from regions with smaller curvature radii, whereas lower-frequency bursts (e.g., Cluster 4) may arise from regions with larger curvature radii.
All burst clusters exhibit narrow spectral features, with bandwidth ratios smaller than 0.4. Such narrow spectra may arise from coherent emission by quasiperiodically distributed bunches, where the spectral width depends on the spatial distribution and separation of the emitting bunches \citep{2023MNRAS.522.4907Y,2024A&A...685A..87W}. 
Alternatively, the weakly magnetized synchrotron maser mechanism can naturally produce narrow spectra, with differences in the central frequency potentially arising from local plasma-density inhomogeneities \citep{2025A&A...695A.100L,2025A&A...702A..30L}.

In addition to possible differences in the intrinsic emission processes, propagation effects may also contribute to the observed diversity in the spectral properties among different clusters. FRB 20121102A is embedded in an extreme and dynamic magneto-ionic environment and is spatially associated with a compact persistent radio source \citep{2017Natur.541...58C,2018Natur.553..182M}. Plasma lensing caused by inhomogeneous electron-density structures along the line of sight can introduce frequency-dependent magnification, potentially modifying the observed spectral peak and producing narrow-band spectral features \citep{2017ApJ...842...35C}. Furthermore, free-free absorption in dense plasma environments preferentially suppresses low-frequency emission, potentially leading to an apparent increase in the observed central frequency and a reduction of the effective bandwidth. On the other hand, induced Compton scattering can modify the frequency distribution of FRB radiation during propagation through dense plasma, and subsequent absorption processes may suppress low-frequency emission, thereby affecting the observed spectral properties \citep{2019MNRAS.485.4091M}.

\section{Conclusions}\label{sec5}
In this paper, we have performed an unsupervised machine learning analysis of bursts from three active episodes of FRB 20121102A observed by FAST. Specifically, we use UMAP for dimensionality reduction and HDBSCAN to identify burst clusters in the embedding space. We use width, bandwidth, bandwidth ratio, central frequency, peak flux, fluence, energy, and waiting time as input features, and optimize the UMAP and HDBSCAN hyperparameters based on the Silhouette Score. Our main conclusions are summarized as follows:
\begin{enumerate}
\item Bursts from different active episodes exhibit different distributions in the UMAP embedding space, where HDBSCAN identifies four well-separated clusters with noise. The clustering structure remains broadly stable across the tested ranges of UMAP and HDBSCAN hyperparameters. The  
burst clusters are primarily distinguished by spectral properties, including central frequency, bandwidth, and bandwidth ratio. MI analysis further indicates that these parameters are strongly associated with the UMAP embedding and HDBSCAN clustering results, highlighting the importance of spectral properties in revealing the complex substructures of repeating FRBs. These spectral differences may be related to the geometry or local physical conditions of the emission regions.

\item In addition to its distinct spectral properties, Cluster 1 exhibits a significantly different energy distribution from the other clusters. It contains almost all high-energy bursts, with an average energy approximately one order of magnitude higher than those of the other clusters. Moreover, Cluster 1 occupies a distinct region in width--fluence parameter space compared with the other clusters. These results suggest that Cluster 1 may represent a physically distinct burst population, potentially associated with different emission conditions or radiation mechanisms.

\item The fractions of different burst clusters show clear variations across active episodes. Cluster 1 accounts for about 27\% of the bursts in 2019, but only about 8\% in 2022 and 2023. Cluster 2 has the largest fraction in 2019 and 2022, with its bursts characterized by higher central frequencies and narrower bandwidths, while Cluster 3 has the largest fraction in 2023, with its bursts characterized by broader bandwidths and intermediate central frequencies.
These variations may indicate changes in the physical conditions of the emission regions or the local propagation environment across active episodes.
\end{enumerate}

\begin{acknowledgments}
We thank the anonymous referee for helpful and insightful comments that improved the quality of the manuscript.
We sincerely thank Dr. Yi-Dan Wang and Prof. Pei Wang (National Astronomical Observatories, Chinese Academy of Sciences) for assistance with data collection and helpful discussions. We also thank Xue-Guang Zhang, Ying Gu, Yun-Peng Li, Zi-Qi Wang and Ren-jie Xiong for helpful discussions. 
This work is supported by the National Key R\&D Program (No. 2024YFA161100) , the Guangxi Science and Technology Innovation Platform Program (Leitai Action Plan, Grant No. Guike LT2600640026), Guangxi Key R\&D Program (Guangxi Funeng Action Plan, Grant No. FN2504240030), National Natural Science Foundation of China (grant Nos. 12133003),  and the Innovation Project of Guangxi Graduate Education (grant No. YCBZ2026029). E. W. L. is also supported by the Guangxi Talent Program (``Highland of Innovation Talents'').
\end{acknowledgments}

\appendix
\section{Robustness of Hyperparameter Optimization}\label{app1}
\begin{figure*}[ht!]
\centering
\includegraphics[width=0.48\textwidth]{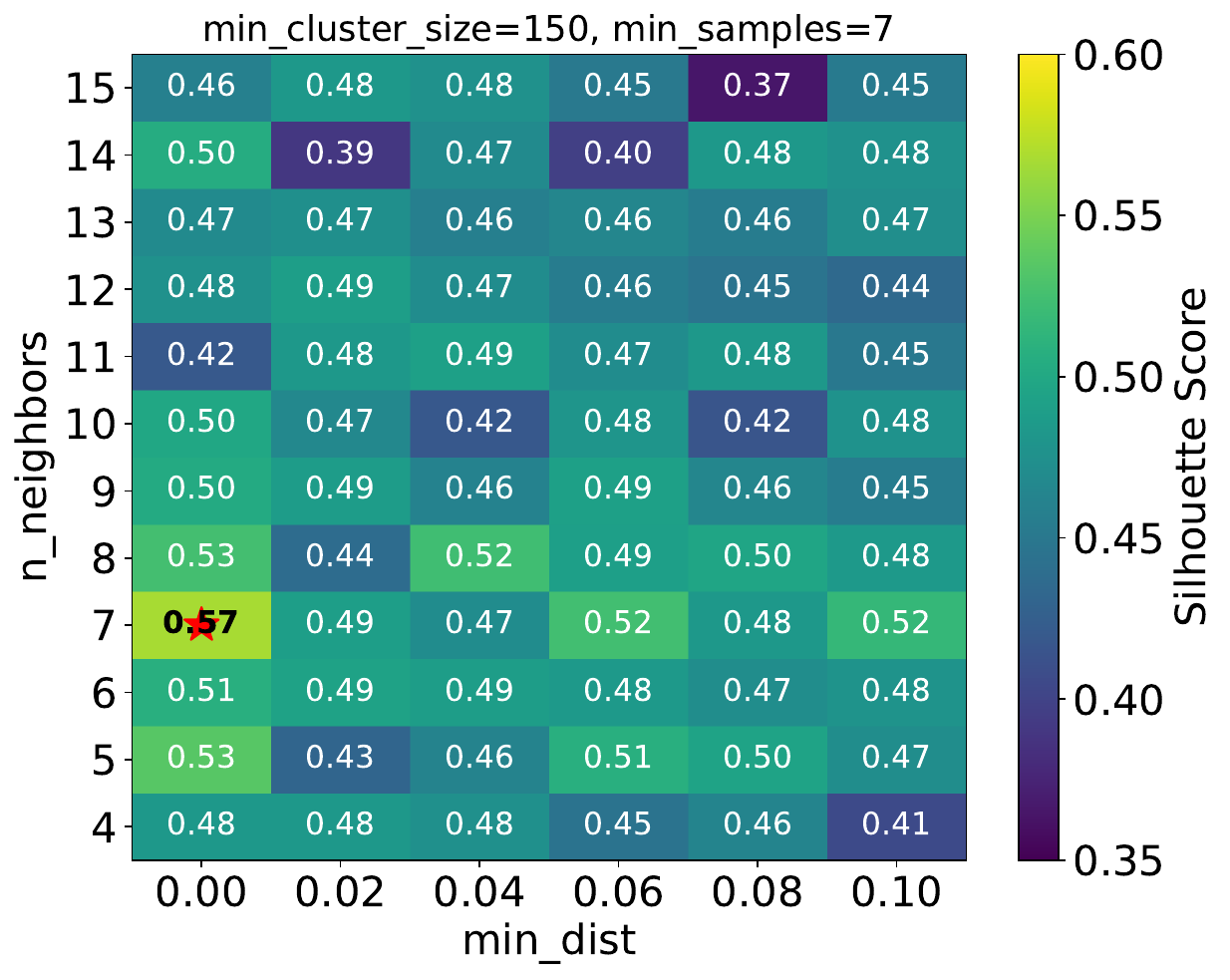}
\includegraphics[width=0.48\textwidth]{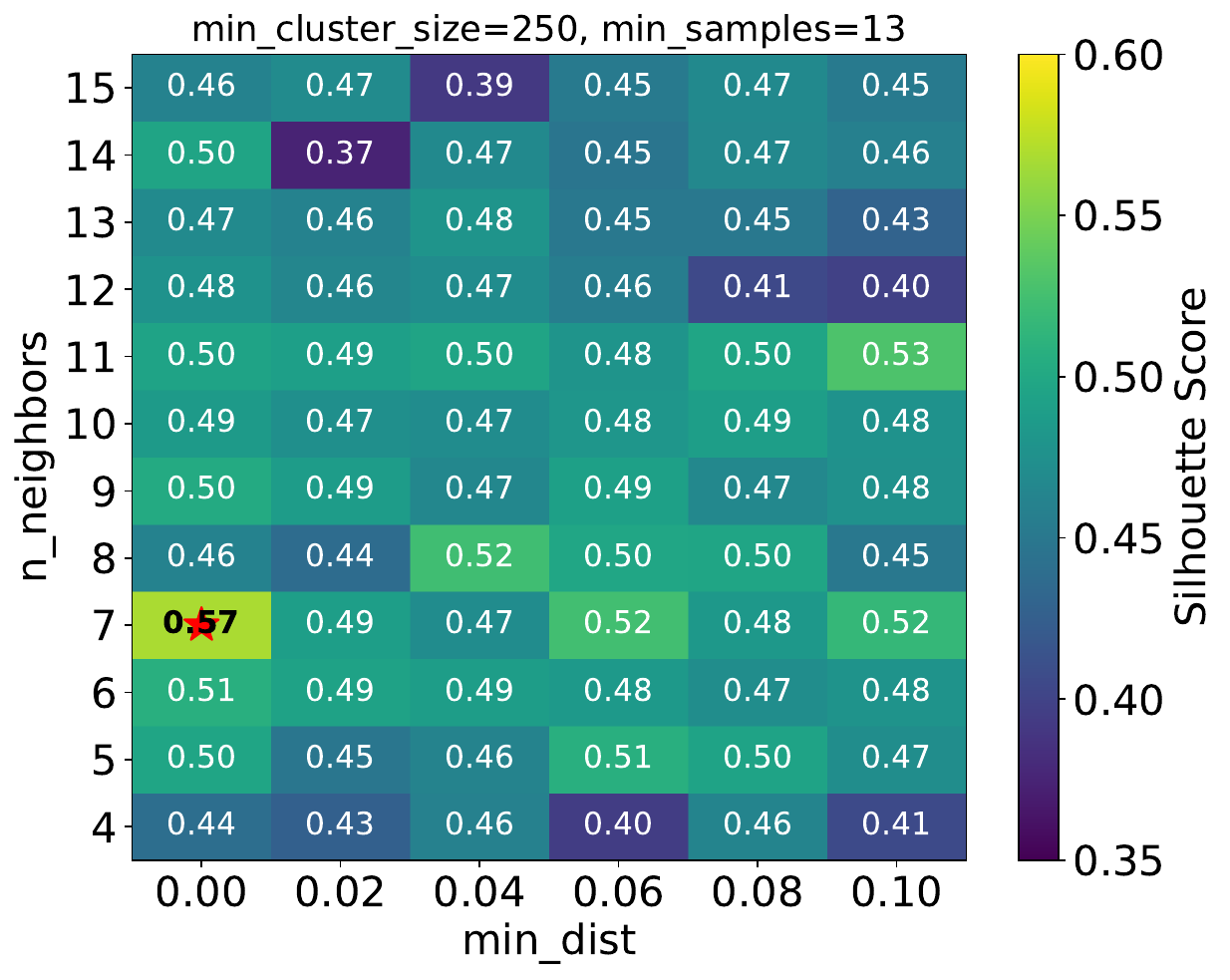}
\includegraphics[width=0.48\textwidth]{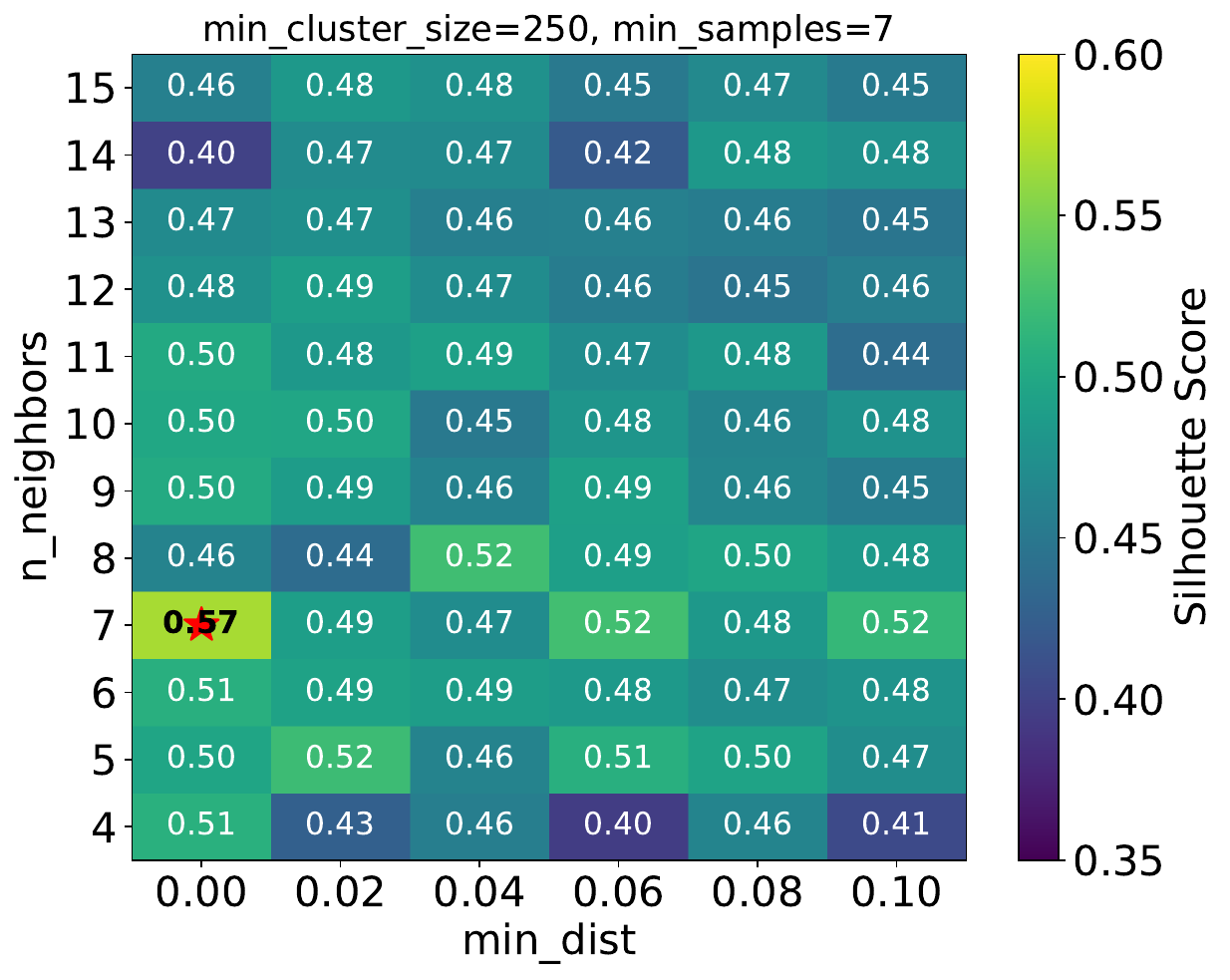}
\includegraphics[width=0.48\textwidth]{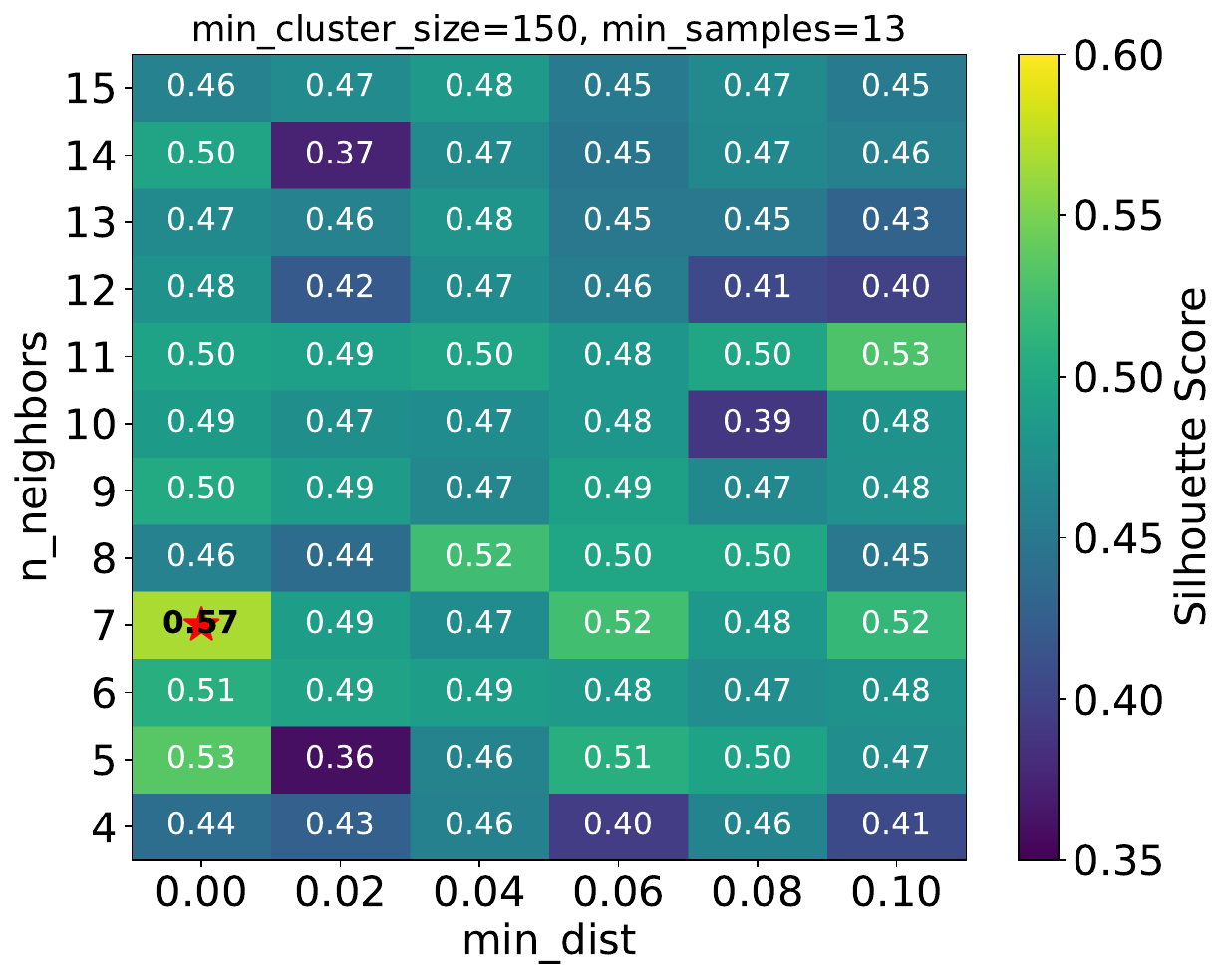}
\caption{Same as the upper panel of Figure \ref{Fig1}, but for different HDBSCAN hyperparameter combinations. The upper-left, lower-left, upper-right, and lower-right panels show results for (\texttt{min\_cluster\_size}, \texttt{min\_samples})=(150, 7), (250, 13), (250, 7), and (150, 13), respectively.}\label{Fig11}
\end{figure*}

To further assess the robustness of the selected UMAP configuration, we present in Figure \ref{Fig11} the Silhouette Scores for different combinations of UMAP hyperparameters under various HDBSCAN parameter settings. The results show that the UMAP configuration with \texttt{n\_neighbors}=7 and \texttt{min\_dist}=0.0 consistently yields the highest Silhouette Score across different HDBSCAN configurations, supporting the robustness of the selected UMAP hyperparameters.

\section{Cluster Properties for an Alternative UMAP Configuration }\label{app2}
\begin{figure*}[ht!]
\centering
\includegraphics[width=\textwidth,keepaspectratio]{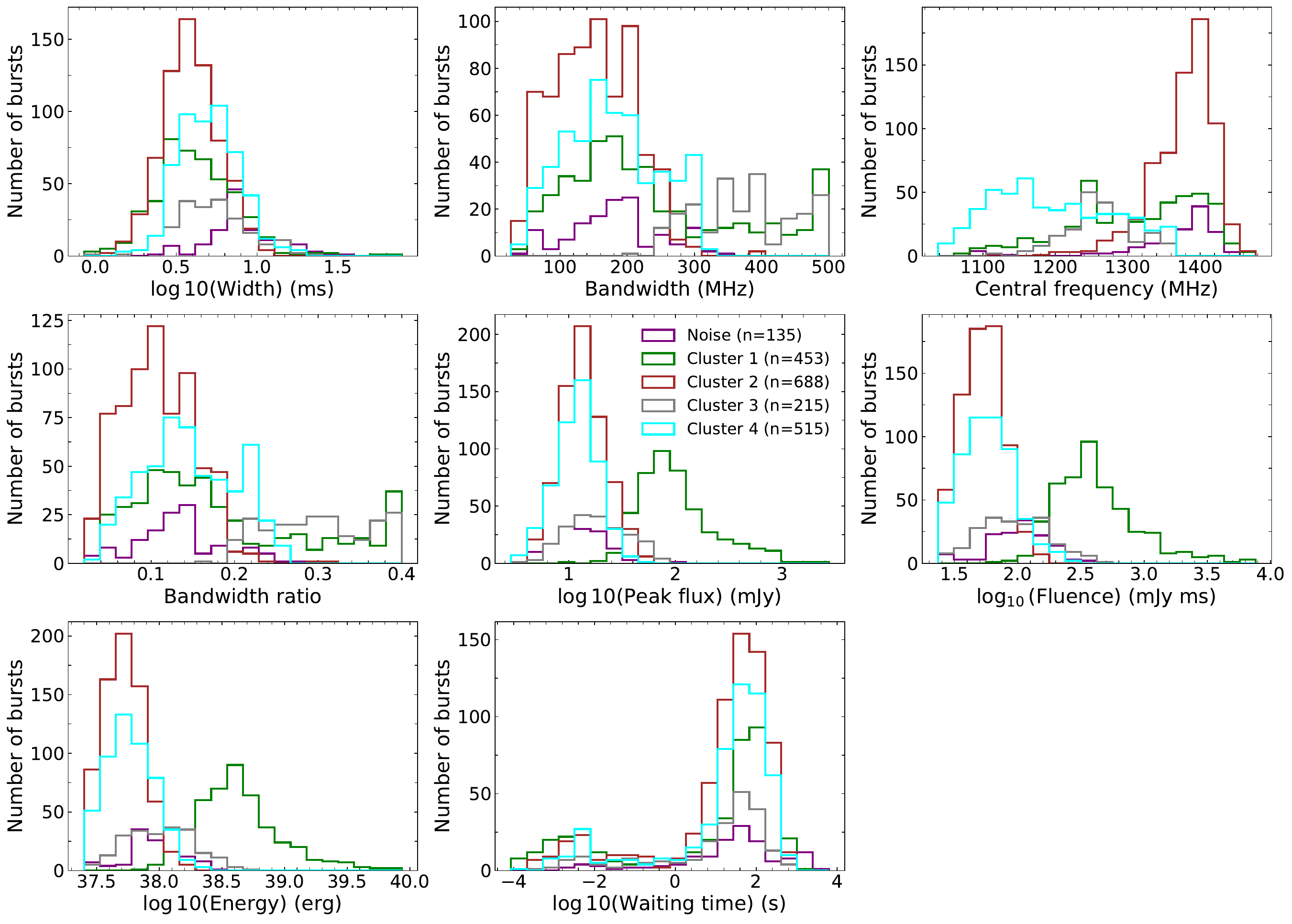}
\caption{Same as Figure \ref{Fig3}, but with \texttt{n\_neighbors}=5, \texttt{min\_dist}=0.0, \texttt{min\_cluster\_size}=200, and \texttt{min\_samples}=10.}\label{Fig12}
\end{figure*}

\begin{figure}[ht!]
\centering
\includegraphics[width=0.45\textwidth]{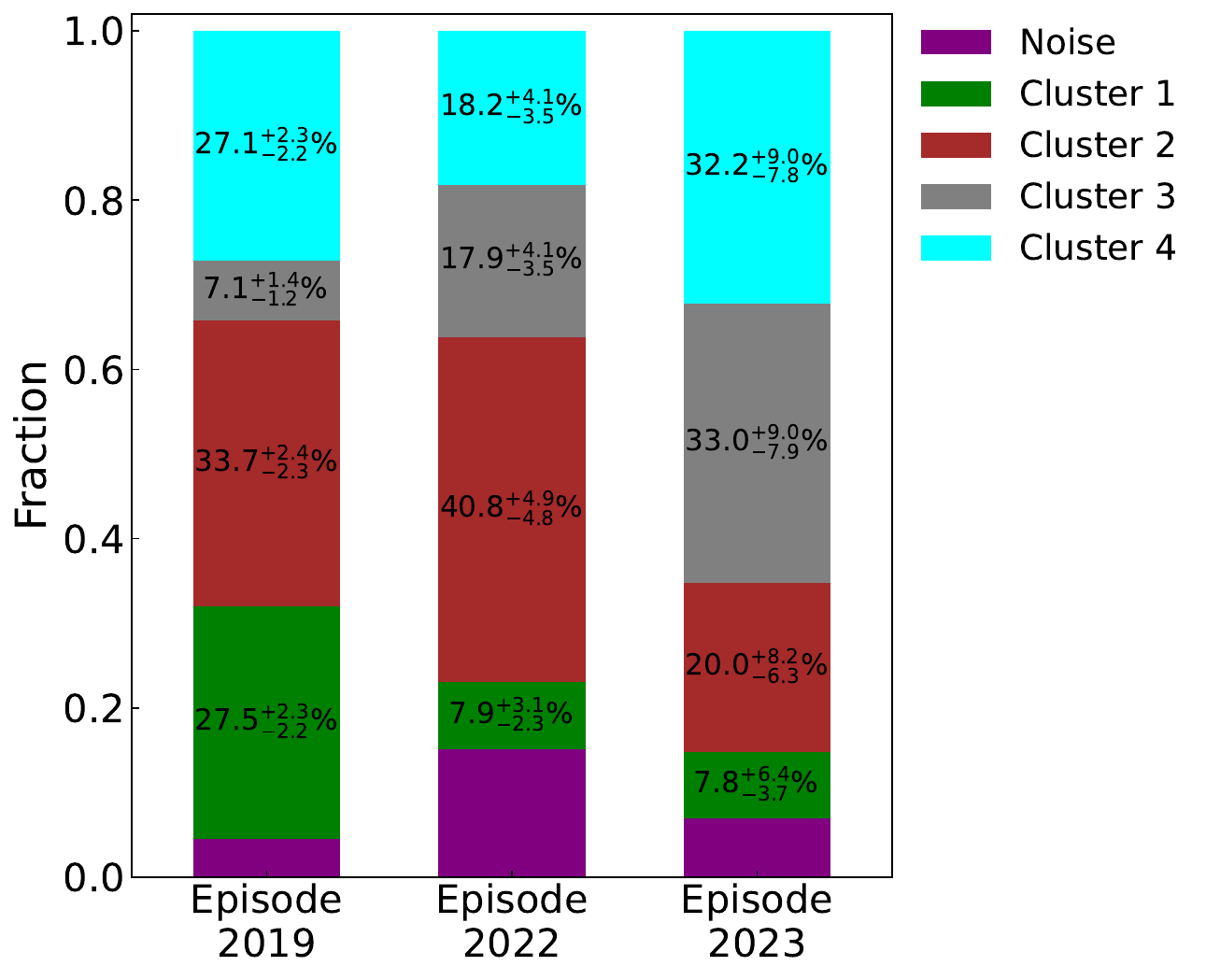}
\caption{Same as Figure \ref{Fig4}, but with \texttt{n\_neighbors}=5, \texttt{min\_dist}=0.0, \texttt{min\_cluster\_size}=200, and \texttt{min\_samples}=10.}\label{Fig13}
\end{figure}

We present the cluster properties obtained with \texttt{n\_neighbors}=5, \texttt{min\_dist}=0.0, \texttt{min\_cluster\_size}=200, and \texttt{min\_samples}=10. Figure \ref{Fig12} displays the distributions of the input features for the identified clusters, while Figure \ref{Fig13} shows the fractions of bursts in each cluster across three active episodes. The cluster properties and the fractions of bursts in each cluster are broadly consistent with those obtained using the adopted configuration.

\clearpage

\bibliography{sample701}
\bibliographystyle{aasjournalv7}

\end{document}